\documentstyle[12pt,twoside,fleqn,epsfig]{article}
\newcommand{\be}{\begin{eqnarray}}
\newcommand{\ee}{\end{eqnarray}}
 
\begin{document}
 
\begin{tabbing}
\`SUNY-NTG-96-16\\
\
\end{tabbing}
\setlength{\parskip}{20pt}
\vbox to  0.8in{}

\centerline{\Large\bf Dilepton/photon production in heavy  ion collisions,}
\centerline{\Large\bf and the QCD phase transition   }
\vskip 2cm

\centerline{\large
  C.M. Hung\footnote{Email: cmhung@insti.physics.sunysb.edu}  and
  E.V. Shuryak \footnote{Email: shuryak@dau.physics.sunysb.edu}}
\vskip .3cm

\centerline{Department of Physics}
\centerline{State University of New York at Stony Brook}
\centerline{Stony Brook, New York 11794}
\vskip 0.35in
\centerline{\bf Abstract}
  We calculate electromagnetic production from highly excited hadronic matter
created in heavy ion collisions. The rates include 
the usual lowest order processes
in quark-gluon plasma plus the usual reactions in the hadronic phase,
 related with $\rho,a_1$ mesons. The space-time integration is done using a
hydrodynamical model. We have found that dilepton mass spectrum agrees with
results of other previous works, but $disagrees$ with the CERES dilepton data.     
In order to explain these data, some ``unconventional" production
mechanism need to be
incorporated: we discuss especially the notion of modified $\rho,a_1$ masses,
which indeed may explain the data. Other suggestions (e.g. the longer-lived
fireball) to increase the production of low mass dileptons seem  to be
insufficient for the task.
The results for direct photon production  are  below the
current WA80 experimental bounds,
 for all variants considered.

\indent
\section{Introduction}

   The main goal of heavy-ion collision program in the
   AGS/SPS energy range (10-200 GeV/A) is to produce hot/dense
hadronic matter with the energy density of the order of few $GeV/fm^3$
and to study its properties. Especially interesting are $early$ stages
of the collisions, when
theory predicts existence of the QCD phase
transition into a new phase, called the Quark-Gluon Plasma (QGP).
However, so far no direct experimental evidence of the 
QGP has been found.
   The main reason for this is well known: strong collective interaction in the
   system as it expands and cools erases most of the traces of the
   dense stage. As a result,  the observed hadrons come mostly from
   a dilute freeze-out stage, with a $final$ temperature
   $T_f=120-140MeV$.\cite{pbmjs}

   One possible way to study the earlier
 stages (to be discussed in this paper) is to look   for  phenomena which
mostly happen very soon after the collision, such as production of
$dileptons$ and $photons$\footnote{Another possibility is to look
   for signals which are $accumulated$ during the  evolution:
the well known examples include excessive production of  
strangeness  or charmonium suppression.}\cite{Shu_78}.
 At high (RHIC/LHC) energies
one may hope for some kinematic
enhancement of the QGP signals because
 the initial stage is much hotter
than hadronic mater, $T_i >> T_c$
 \cite{Shu_hotglue,Geiger}, but this is certainly $not$ the
case for AGS/SPS energies. Therefore, in this energy domain the main
signals
for the phase transition still come from the $hadronic$ stage, not the QGP.

    Experiments designed to observe the {\it direct
   photons} or {\it dilepton continuum} produced by excited hadronic matter
 in heavy ion collisions are generally much more 
difficult to perform compared with  measurements of hadronic observables. 
Therefore, only recently were the first photon and dilepton measurements
announced by four CERN SPS 
experiments:  NA34/3, NA38 and CERES (NA45) for dileptons and WA80 for photons.
It was found that dilepton production exceeds
 backgrounds expected from hadronic 
and charm decays. Furthermore, the signal also exceeds
 theoretical expectations for ``conventional'' processes,
both in hadronic and quark-gluon matter \cite{SX_where}.
 Especially dramatic is the excess observed by CERES \cite{CERES} 
   in the mass region  $M_{e^+e^-}=0.3-0.6$ GeV.
 This observation  has since created
 a rapidly growing 
  theoretical literature.

  Calculation of 
    the dilepton/photon yield consists of two components: (i)
evaluation of production $rates$ (see section \ref{sec_rates}); and (ii)
their integration  over the {\it space-time evolution} of the collision
(see section \ref{sec_hydro}).
 In the so called   
 $conservative$ approach (the well-known
hadronic processes with vacuum parameters and the usual space-time
evolution
of heavy-ion collisions) several groups have obtained rather similar results,
which however do not 
explain the CERES data, neither in magnitude nor even in the shape of the
mass spectrum.

  This situation has lead to many ``unconventional'' hypotheses, which 
 include
(in a more or less
chronological order): (i) dropping $m_\rho$
   \cite{Hof_94,LKB_95}; 
(ii) high pion occupation numbers at low momenta \cite{KKP_93}; 
(iii) a very long-lived fireball \cite{SX_where}; 
(iv) dropping  $m_{\eta'}$ \cite{Shu_94,KKM_96,Wang_96}; (v) a  modified pion
   dispersion curve \cite{Song_96}; (vi) dropping $m_{a_1}$ (discussed below).

Let us start with the possibility (ii), which is usually taken into
account
by introduction of the pion chemical potential $\mu_\pi$, which is
approximately equal
to $m_\pi$. Although the true nature of the low-$p_t$ pions is
not yet completely clear, most probably they  come from
the resonance decay and/or spectral  modification due to collective
potentials. Both are late-stage phenomena, which can hardly affect
the early-stage dilepton production. Furthermore, studies of this
explanation
made in \cite{LKB_95} have shown, that the low-M dilepton enhancement
due to $\mu_\pi\approx m_\pi$ is way too small
 compared to CERES data.
 
  In section \ref{sec_modified} we will look at (i), the T-dependent  $m_\rho$,
and (similar to  \cite{LKB_95}) conclude that it may indeed describe the
data.
Furthermore, we have linked it to (vi) by theoretical
arguments related to chiral symmetry restoration and have derived
experimental consequences of
the most plausible scenario, both for dilepton and photon (section
\ref{sec_photons} ) yields.

  We also study separately a proposal (iii). 
In our previous paper \cite{HS_95}, we found that (at least in
a hydro approach)
the so called ``softest point'' of the Equation of State (EOS)
leads to especially long-lived fireball. Although it is expected to
happen
 at collision energies
way below those for CERN experiments, in
section \ref{sec_long} we have pushed this idea to the extreme and assumed the
scenario with the long-lived fireball. The results for the dileptons  
are quite disappointing: very different scenarios of
space-time evolutions give very similar dilepton mass spectra.
So, with 
the standard dilepton rates and fixed masses it is not possible
to explain the observed dilepton yield by longer lifetime. 

\section{Electromagnetic processes in hadronic matter}
\label{sec_rates}

    Dilepton/photon production
in the QGP phase is based on fundamental QCD processes like $\bar
q q \rightarrow e^+ e^-$   \cite{Shu_78} and was calculated long ago.
The rate in the pion gas due to $\pi\pi$ annihilation was considered
in \cite{Kapusta_etal}, and those two basic processes can be included by
 the ``standard rate'' formula:
\begin{eqnarray}
{dR\over d^4q} = {\alpha^2 \over 48\pi^4} F e^{-{q_0\over T}}
\end{eqnarray}
where the rate $R$ is counted per unit volume per unit time, $q$ is 4-momentum of
the virtual photon ($q^2=M^2_{e^+e^-}=M^2$),
$F$ is a constant in QGP and 
the usual pion form-factor in the pion gas,
which can be written in standard vector-dominance form
\footnote{For the quark masses, we adopt the perturbative result \cite{Kapusta_etal}
$m_q(T) = {gT\over \sqrt{6}}$, with $g = 2.05$ which corresponds to
$\alpha_s = 0.33$.}
\begin{eqnarray}
\label{ff}
F = \cases{F_H \buildrel \rm def \over = {m_\rho^4\over[(m_\rho^2 -
    M^2)^2 + m_\rho^2\Gamma_\rho^2]}, &(Hadronic)\cr 
F_Q \buildrel \rm def \over = 12 \sum_q e^2_q\left(1+{2m_q^2\over
  M^2}\right) \left(1-{4m_q^2\over M^2}\right)^{1\over 2}, &(QGP)\cr}
\end{eqnarray}
Later  additional processes including $\rho$ mesons were added.  
It was also pointed out in \cite{XSB} that $a_1$ meson is very
important\footnote{
In order to explain why $a_1$ is important, let us go ``backward in time'':
it is the first hadronic resonance which
may be excited in a collision of a photon, real or virtual, with a pion.
},
especially for photons and low-mass dileptons. Further work was done
in refs. 
 \cite{Song_93,SKG_94}. 

  Before we come to specific formulae, let us remark on some
  misunderstanding of the role of the resonances, which has 
even led to double counting in some previous papers.
 Unstable particles like $\rho$ meson can
be considered either (i) as parents of decay processes (
such as
$\rho\rightarrow  e^+ e^-, a_1 \rightarrow  e^+ e^- \pi$ )
 or (ii) as intermediate states in
particular reactions with ``stable'' particles
(e.g. $\pi^+\pi^-\rightarrow  e^+ e^-$). Furthermore, the same
resonance can enter as an intermediate stage in many different reactions: This
created the impression that by considering many of these reactions, one can in fact
increase the dilepton yield.

  However, this is not true.
 In {\it thermal equilibrium} the average number of
mesons depends on their mass but should $not$ depend on their width
(which only shows how often these particles are created and decay).
In order to understand  how this happens (we'll give
detailed discussion in the Appendix),
recall that standard Breit-Wigner amplitude
 is proportional to $\Gamma_{in} \Gamma_{out}/((M-M_{res})^2
 +\Gamma^2_{tot}/4)$. Summing over all possible ``in'' channel one
 gets
$\Gamma_{tot}$ in the numerator, and after that one may
approximately\footnote{Provided the total width is not large compared with
  the temperature!}
 substitute the Breit-Wigner amplitude
simply
 by $
\Gamma_{out} \delta(M-M_{res})$. The resulting rate is nothing else but the decay 
contribution (i) mentioned above: clearly one should not include it twice.
\begin{figure}[!h]
\begin{center}
\includegraphics[width=6.cm]{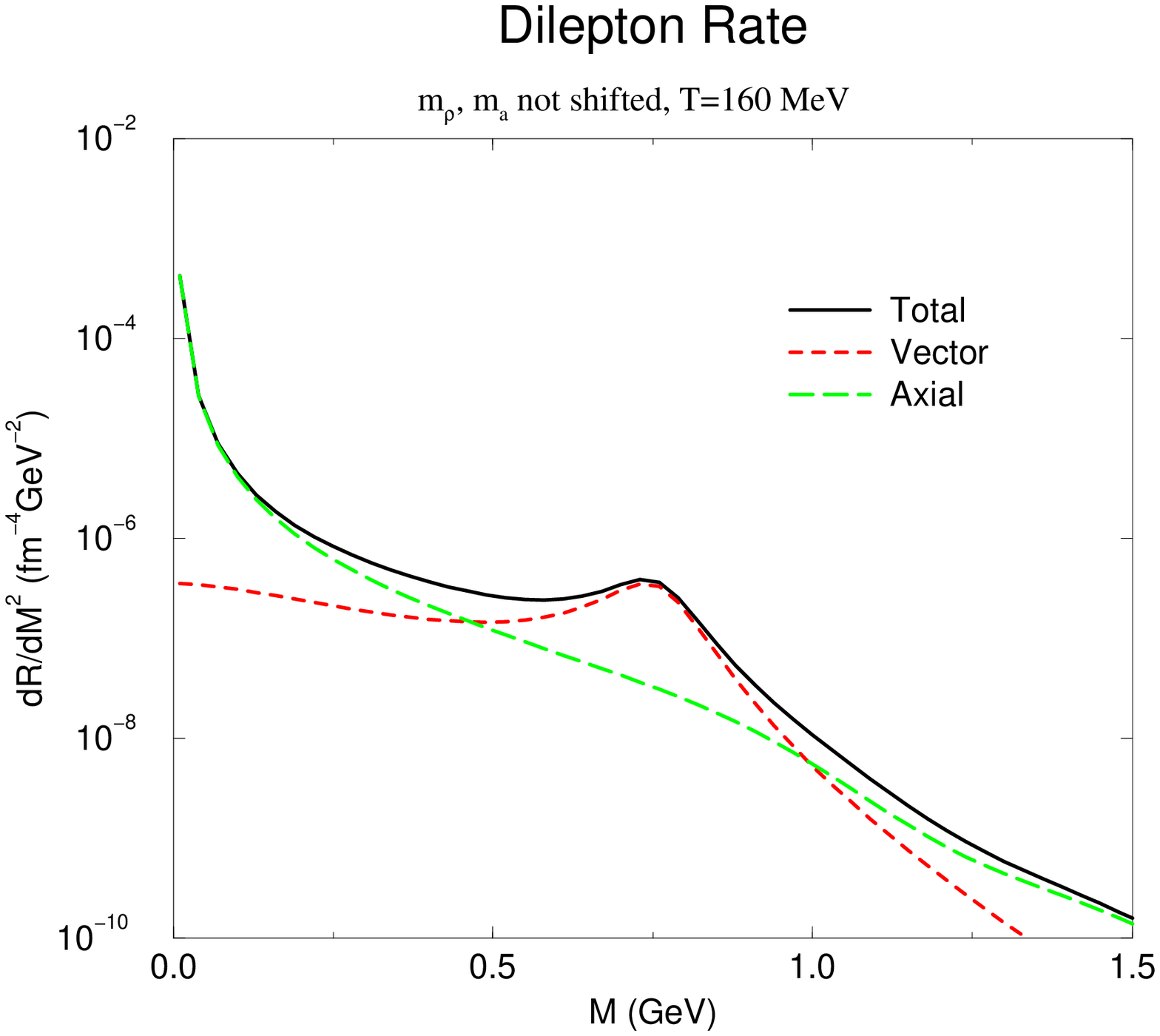}
\hspace{.2cm}
\includegraphics[width=6.cm]{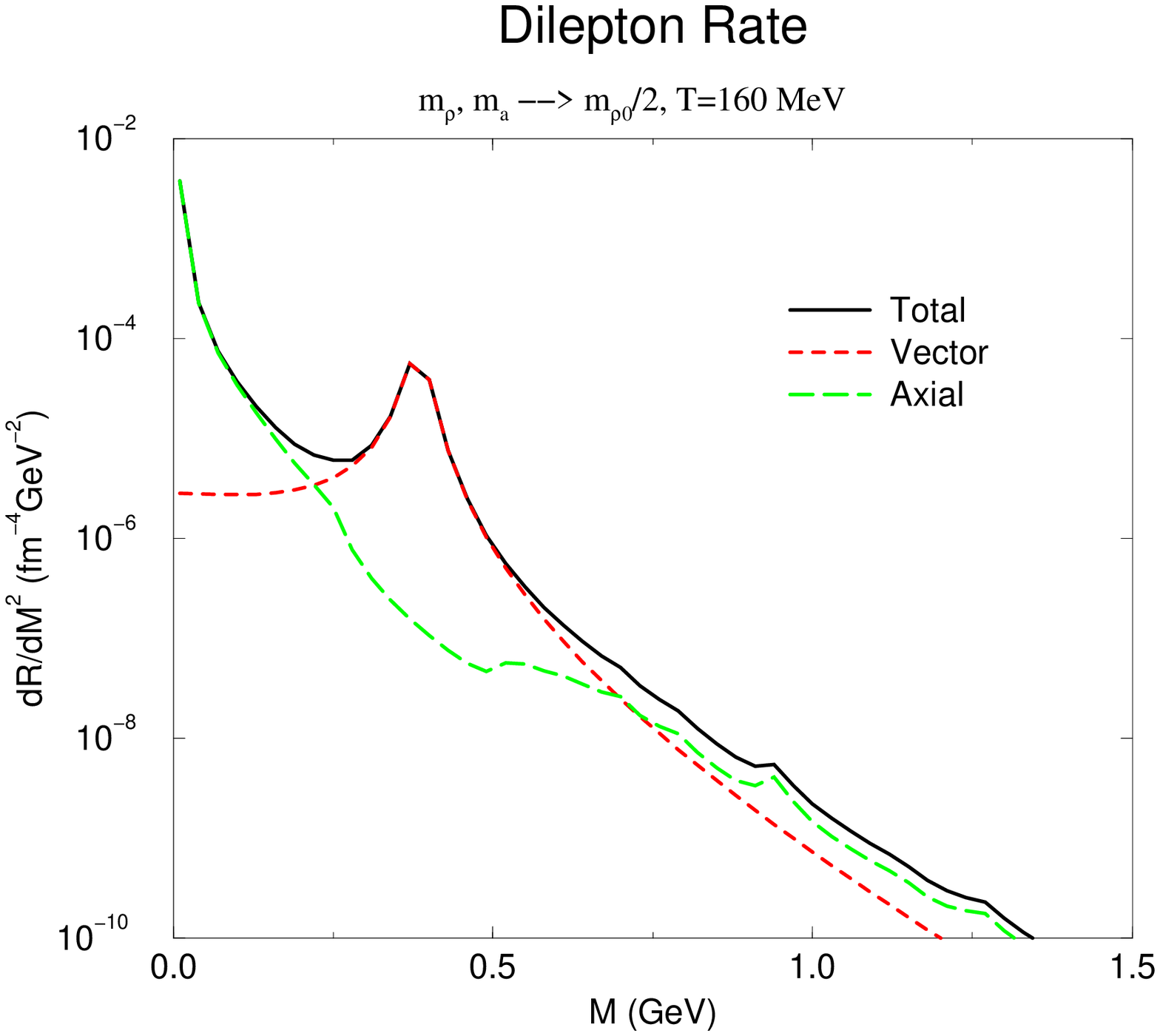}
\end{center}
\caption{\label{rates}
Dilepton production rates
(a) standard (b) comparison
with maximally shifted according to variant D discussed in section
\ref{sec_modified}. 
}\end{figure}
We have found it rather  convenient for our applications to use
expressions recently derived in \cite{SYZ_96}: They contain
processes in zeroth and first order in the pion density.
The expression for the dilepton rate reads
\be
\label{jimrate}
\nonumber {dR \over d^4q} = &&{\alpha^2 \over 3 \pi^3 q^2 }{ 1 \over 1+  exp(q_0/T)}
\left[ 3q^2 Im \Pi_v(q^2) \right. + \int {d^3k
  n(k_0)\over (2\pi)^3 2k_0 f_\pi^2} \\
\nonumber && [-12 q^2 Im \Pi_v(q^2) + 6(k+q)^2
  Im \Pi_a((k+q)^2) \\ 
&& \left. + 6(k-q)^2
  Im \Pi_a((k-q)^2)] \right]
\ee
where k is the pion and q is the virtual photon momentum. 
Here $Im \Pi_v, Im \Pi_a$ are imaginary parts (or spectral densities)
for vector and axial currents. If this expression is understood as
expansion in the pion density, they should be evaluated in $vacuum$,
and thus related to experimental data on $e^+ e^- \rightarrow hadrons$
and
$\tau$ lepton decay. Furthermore, $Im \Pi_v, Im \Pi_a$
  can be approximated by well-known
contributions of $\rho,a_1$ resonances, which produce dileptons by their decays
 into $e^+ e^-$ and $\pi e^+ e^-$,    respectively. We chose the
 following parametrizations for $Im\Pi_v$ and $Im\Pi_a$ (see section
 \ref{sec_modified})
\be
Im \Pi_v = {2f_\pi^2\over M^2}{m^2_{\rho 0} m_\rho\Gamma_\rho\over (M^2-m_\rho^2)^2+m_\rho^2\Gamma_\rho^2}\\ 
 Im \Pi_a = {f_a^2m_a\Gamma_a\over (M^2-m_a^2)^2 + m_a^2\Gamma_a^2}
\ee
where $f_\pi = $93 MeV, $\Gamma_{\rho} = \Gamma_{\rho 0}{M^2\over
  m^2_{\rho 0}}$, $\Gamma_{\rho 0} =$ 149 MeV, $m_{\rho 0} =$ 770 MeV,
$f_a =$190 MeV, $\Gamma_{a} = \Gamma_{a0}{M^2\over
  m^2_{a0}}$, $\Gamma_{a0} = $ 400 MeV, $m_{a0} = $ 1210 MeV \cite{SYZ_96}.

 The resulting
 rates 
\cite{SYZ_96} are shown 
in Fig.\ref{rates}(a), where we
  show the contributions of the vector and
axial parts separately.  We have used simple Breit-Wigner
parametrization\footnote{
This is why vector contributions does not vanish below $2m_\pi$.} with
$vacuum$
parameters for resonances taken from Particle Data Tables, while part (b)
corresponds
to both $\rho,a_1$ masses shifted to ${1\over 2}m_{\rho 0}$ in the ``mixed phase''.
(We assume the critical temperature  $T_c$=160 $MeV$, and 
the mass shifts  will be discussed in detail below).
In both cases, the direct channel resonance $\rho$  dominates around
its mass, while the $a_1$ contribution takes over at small masses.

\section{The model for the space-time evolution}
\label{sec_hydro}

 The hydrodynamical
model was suggested by Landau more than 40 years ago,
and its applications to heavy ion collisions
has a long history. Experiments with
heavy ions at lower energies ($\sim 1 $ GeV/A) were able to detect
collective motion of nuclear matter by comparing velocity distribution
of different nuclear fragments. 

 One important finding is that in the AGS/SPS energy range for heaviest
nuclei (Au Au at AGS and Pb Pb at SPS)  the 
rapidity spectra of  $\pi,K,p,d$  are consistent
 \cite{pbmjs} with
a simple hydro description: a convolution of
{\it thermal motion} at breakup
 (which depends on the particle mass) with a  collective flow
{\it common for all  species  including baryons}\footnote{In
the framework of  cascade models such as RQMD this topic was
studied \cite{sorge} 
and it was also concluded that if one cut the excited system into
elements, the
 mean velocity of different species are 
the same with accuracy 10-20 percent.}.
 An open question  however is whether thermalization is rapid
enough, so that  one can
use hydro description from very early times,  or  
 collective motion is  formed only at
later
stages.
 In this work we assume that the former is the case, and will use
hydro description  for
dilepton/photon production. 

\begin{figure}[!h]
\begin{center}
\includegraphics[width=8.cm]{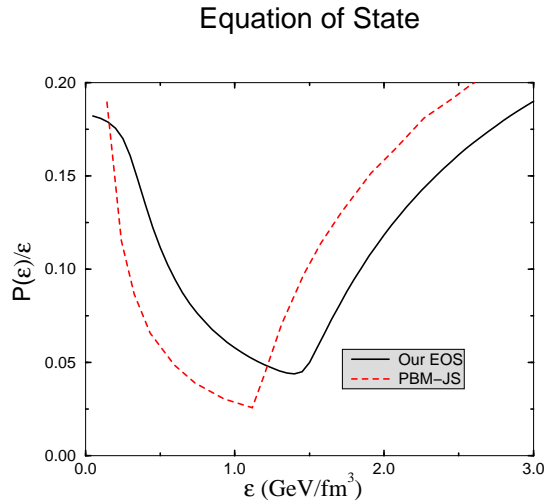}
\end{center}
\caption{\label{eos}The EOS in hydro-relevant coordinates: the ratio of pressure
to the energy density $p/\epsilon$ versus $\epsilon$. Notice
the minimum, which we refer to as the  ``softest point''.  The dashed line
 is the EOS for finite baryon chemical potential $\mu_b=.54$ GeV
from  Braun-Munzinger and Stachel. }
\end{figure}

   The second important observation is related with the equation of
   state
(EOS).
 For finite T and $zero$ baryon density, a conventional
 parametrization  of the  EOS can be given by 
a ``resonance gas" \cite{resonancegas} below $T_c$, plus a bag-model
QGP above $T_c$. We fixed 
 the
phase transition point (smoothened for numerical purposes) at $T_c=160$ MeV.
In hadronic phase the speed of sound 
$dp/d\epsilon=c_s^2 =
0.19$, and in the plasma 
phase the bag constant is $B=0.32\ {\rm GeV/fm^3}$. 
The results are plotted in Fig.\ref{eos}.
This EOS is in reasonable agreement with the lattice results.
  Unfortunately,  there is so far little progress in lattice
simulations for non-zero baryon density. 
People have extrapolated various models which are successful
at nuclear densities, like
Walecka
model, but it is unclear how well such extrapolation works.
Some guidance can probably be provided by hadronic gas including
baryonic resonances, for example the one discussed in \cite{pbmjs}.    
 In  Fig.\ref{eos} we have shown  the corresponding
curve (dashed line) for  baryon chemical potential $\mu_b=.54$ GeV,
corresponding
to AGS breakup conditions. In this case
the baryon/meson ratio is  about 1, one order of magnitude larger 
than baryon admixture
in central region at SPS energies: and
 still the EOS in $p,\epsilon$ coordinates looks
approximately the same as the one we adopted.

   In summary, the first observation shows that for heavy ions
the flow of the entropy (mesons) and the baryonic charge
(nucleons) are about the same. The second shows that EOS (in
$p(\epsilon)$
form) is not much affected by baryonic charge.     
  Taken together, they provide a reason to
 $ignore$ baryonic charge in our hydro calculations.

  We use standard equations of (non-viscous) 
relativistic
hydrodynamics for central collisions (i.e. we assume axial symmetry).
The hydro equations were solved numerically using
the first-order Lax finite 
difference scheme. Energy and entropy 
conservation is monitored, and we have  also made comparison
 with results of several
 earlier works was made to ensure that technical aspects are under control.

   The major uncertainties one faces dealing with hydrodynamical models are 
the initial conditions\footnote{
Although cascade-type event generators (Venus,RQMD or ARC)
 provide some guidance,
 their physical basis is questionable exactly at the first 1-2 fm/c 
of the collision, when we need it.}.
Those can be described by the following set of parameters:
(i) the initial size $z_0$; (ii) the fraction of thermalized
energy/collision energy
$\kappa$; and (iii)  $v_{i,z}$ describing initial distribution of the
longitudinal velocity 
\be
v_z(t=0)=v_{i,z}*tanh(z/z_0) 
\ee
  We consider specifically
 two cases of $central$ collisions: (i)
 S-Au  200 GeV/N and (ii) Pb Au  160 GeV/N for which
the dilepton/photon 
  data were taken.  
  In all cases considered, at t = 0 we assume that $some$ $part$ $\kappa$ of
total energy goes into the thermalized
 matter  at rest, and fix $\kappa$ demanding that at the end of the expansion
(at $T_f=140 MeV$) the predicted number of pions\footnote{
Multiple studies have shown that about 1/3 of pions come from resonance decays.
We included this fact in the normalization.}
 is the same as observed.  

\begin{figure}[!h]
\begin{center}
\includegraphics[width=12.cm]{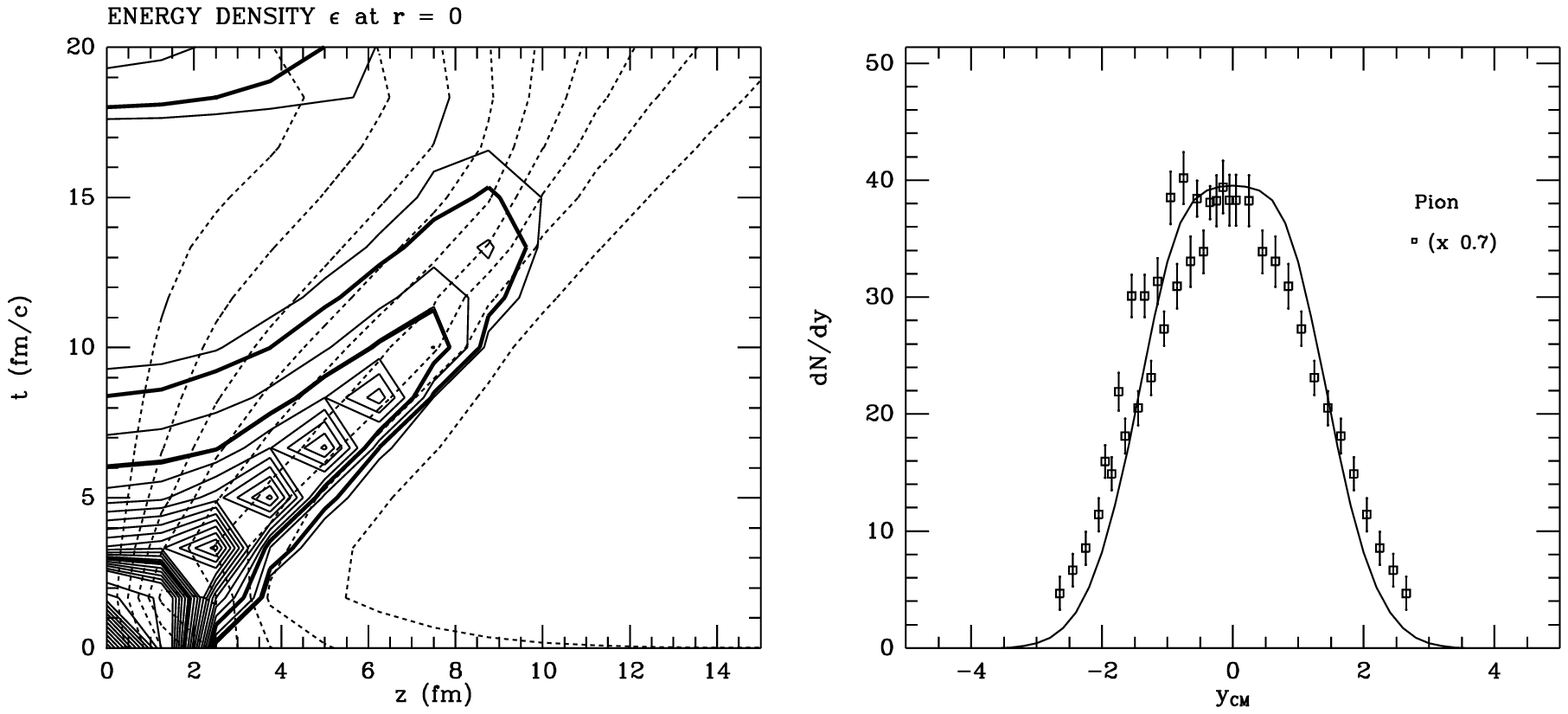}
\includegraphics[width=12.cm]{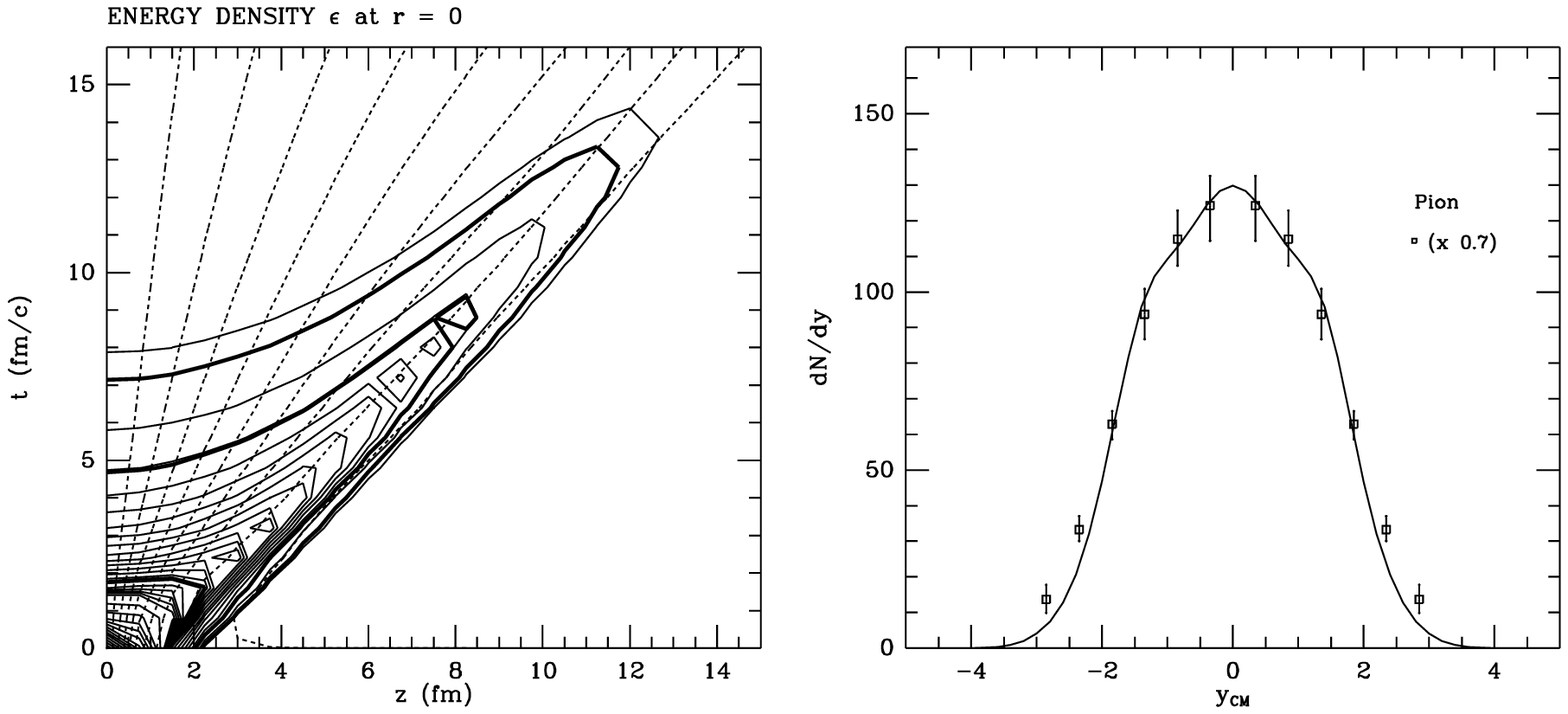}
\end{center}
\caption{\label{hydro_data}
Typical  hydro evolutions for  200A GeV S+Au (upper part)
and  160A GeV Pb+Au (lower part).
The $left$ part shows hydrodynamical solution
in the plane time-longitudinal coordinate. Solid lines are
lines of constant energy density, the dotted ones correspond to
constant longitudinal velocity. Thicker lines show the end of the
mixed phase ($\epsilon = 1.47 \rm GeV/fm^3$) and the break-up
conditions ($\epsilon = 0.31 \rm GeV/fm^3$): at $z = 0$
that happens at about time 5 and 7 fm/c, respectively.
The figures on the right shows the calculated  pion rapidity
spectra (lines) compared with data, where the data has been
scaled by 0.7 to account approximately for resonances.
}\end{figure}

 One extreme scenario is complete equilibration,  $v_{i,z}=0$
\`a la Fermi-Landau. Furthermore, if thermalization is very rapid, the
 usual Lorentz contraction is enhanced 
by the ordinary compression of matter by shocks, leading to very high  
$\epsilon_i\sim 10\ {\rm GeV/fm^3}$ initial energy density,
 well above the phase transition region. 
As shown  e.g. in \cite{marburg} this scenario
is incompatible with SPS data on rapidity distribution.
However, as shown in Fig.\ref{hydro_data}(a), for
   $v_{i,z}= 0.9$ one obtains reasonable description of the
longitudinal motion\footnote{Distribution of other secondaries, as
  well as 
the $transverse$ motion we plan to present elsewhere.  The hardon
rapidity data is taken from \cite{seyboth}
(The data actually refers to
negative hadrons. We scaled them by 0.7 to account for resonance
decays. The $y > 2.65$ ${dN\over dy}$ data for S+Au has been reflected
about $y = 2.65$ to give data for $y < 2.65$.).
}.

\section{Conventional dilepton yields}
\label{sec_conventional}

   Now we evaluate  the dilepton yield, by
 integrating the production rate over the space-time.
The integral is taken between the surface at which initial conditions
are set (the so called ``pre-equilibrium'' contribution is thus left
over) and the ``breakup'' surface, at which the density is so small that
 secondaries can fly away without interaction  
\footnote{In principle, some pion annihilation 
  can happen even later: we have evaluated this
  contribution and have found that it is small. Note also that $\rho,\omega,\eta$
etc decays after breakup are precisely the ``hadronic background'',
which was separately calculated, by the experimental group itself.}.
The surface should be determined from the mean free path, and for pions
it corresponds to $T_{breakup}= 140 MeV$.
Furthermore, in order to compare with CERES data one has to apply their
experimental cuts: For S+Au, $p_T > 200$ MeV/c; $\Theta_{ee} > 35$ mrad; $2.1 < \eta
< 2.65$.
For Pb+Au, $p_T > 175$ MeV/c; $\Theta_{ee} >$ 35 mrad; $2.1 < \eta
< 2.65$.

\begin{figure}[!h]
\begin{center}
\includegraphics[width=6.cm]{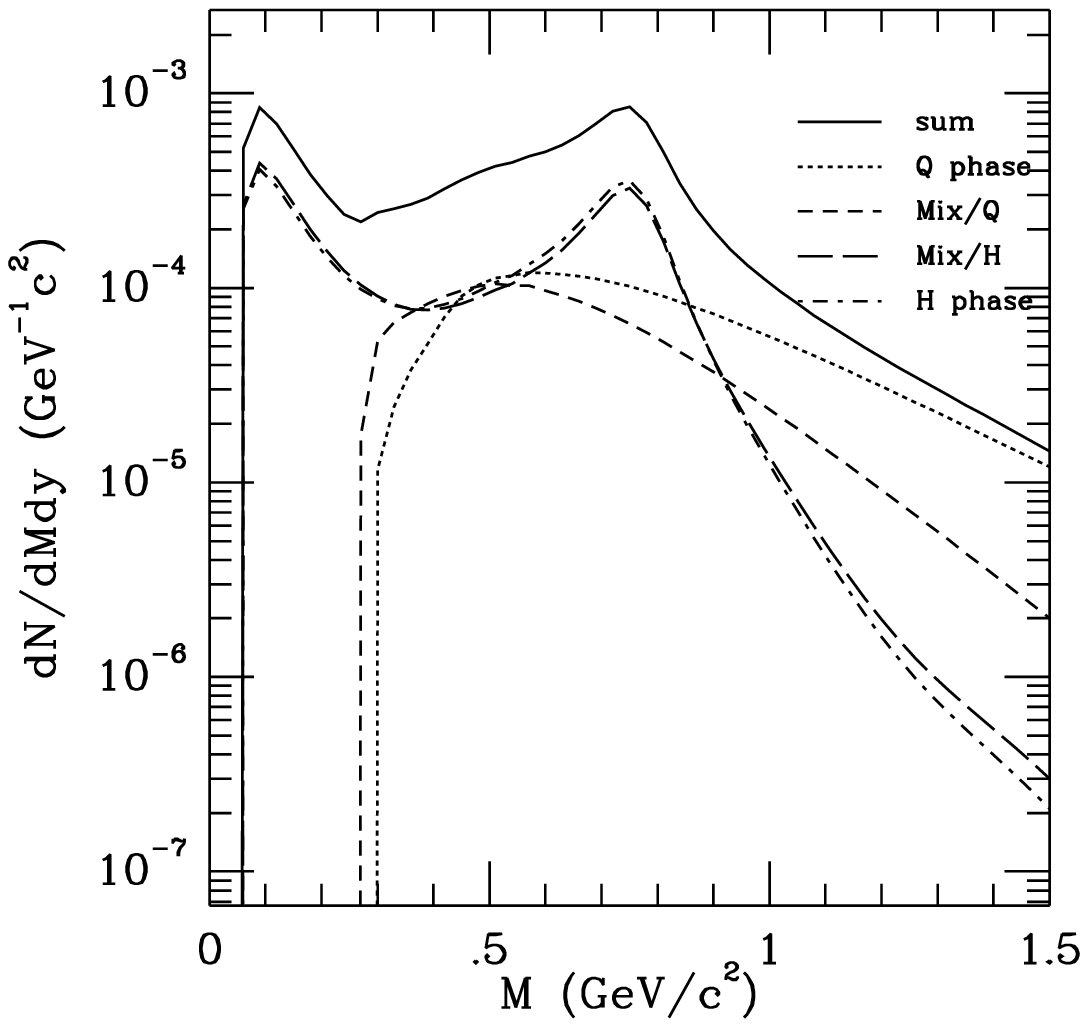}
\hspace{.2cm}
\includegraphics[width=6.cm]{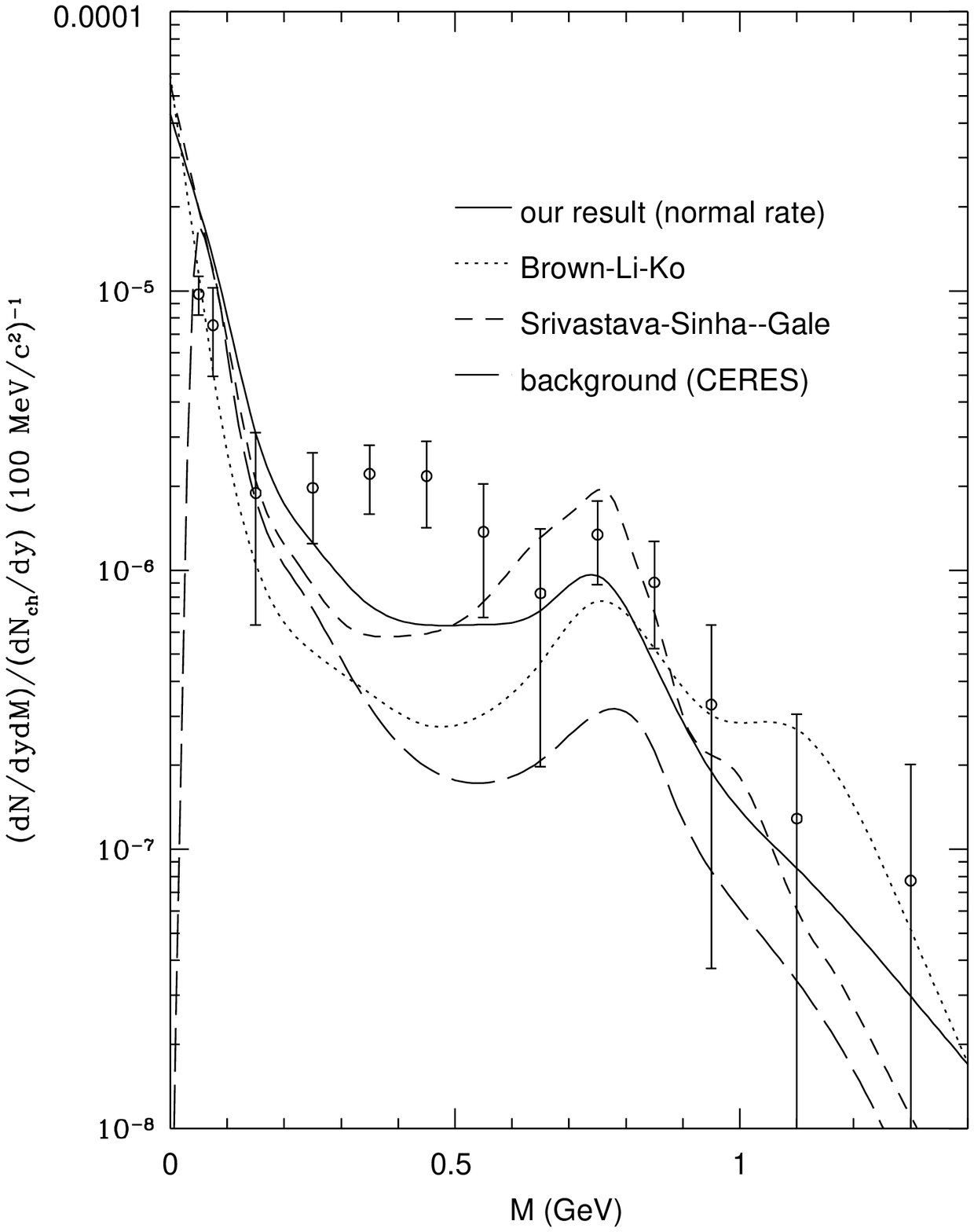}
\hspace{.2cm}
\includegraphics[width=6.cm]{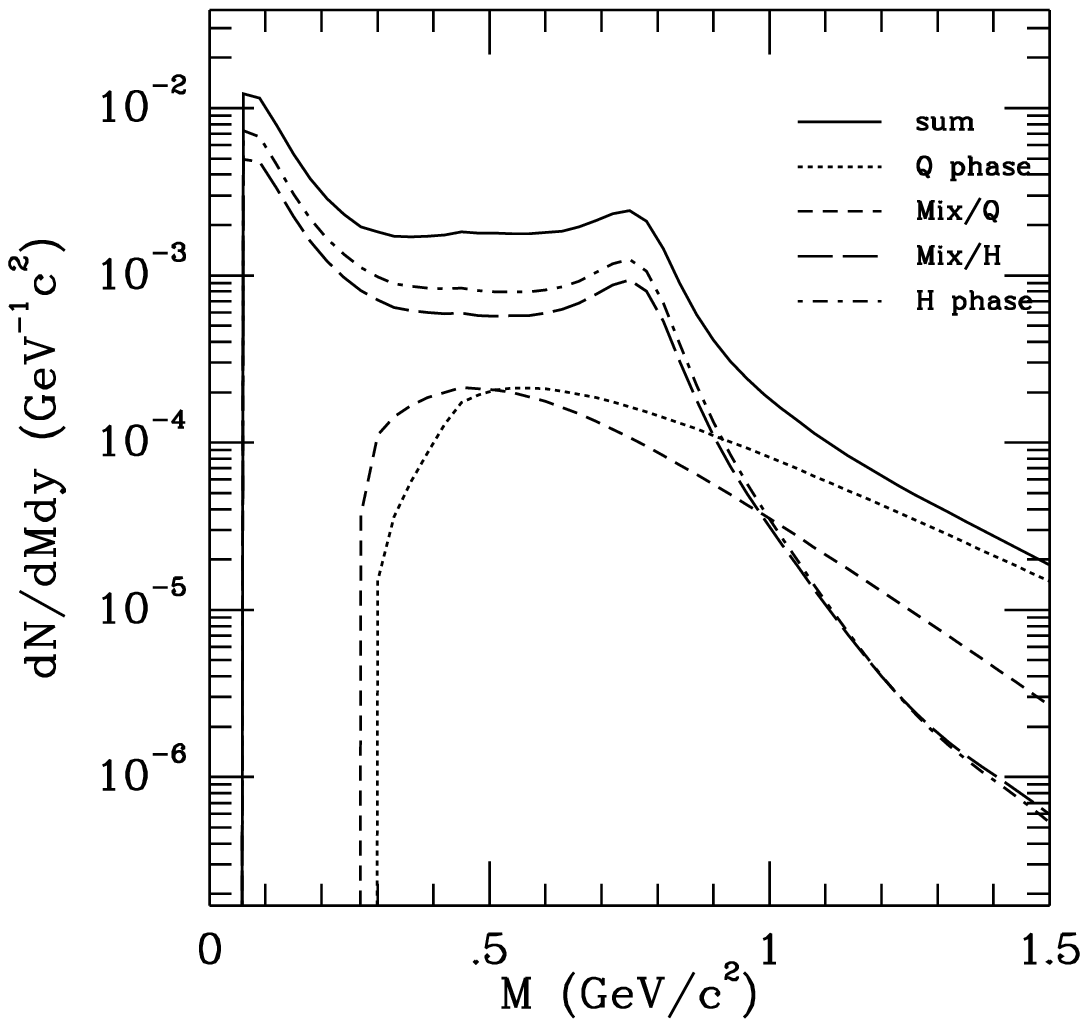}
\hspace{.2cm}
\includegraphics[width=6.cm]{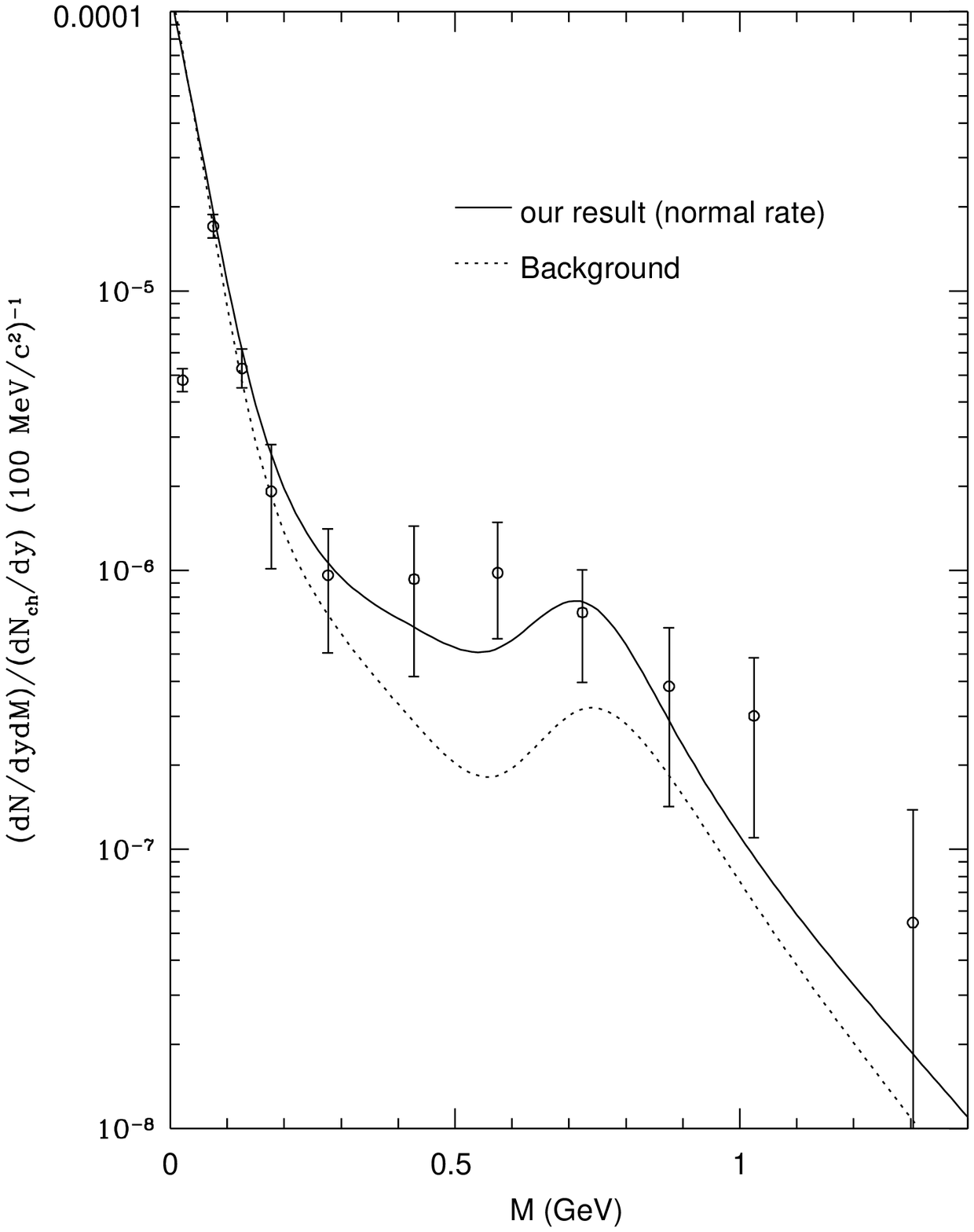}
\end{center}
\caption{\label{Mspectrum_conventional}
CERES dilepton spectra from conventional sources.
(Top) our conventional spectra for 200A GeV S+Au,
shown without and with background from hadronic decays; (Bottom) 
our results for 160A GeV Pb+Au w/o and with background.
}\end{figure}

The results we obtained are shown in Fig.\ref{Mspectrum_conventional}(a):
they are split into 4 contributions:  from pure QGP
 phase (denoted by Q); QGP part of the mixed phase (Mix/Q); hadronic
 part of the mixed phase (Mix/H); and finally from the hadronic phase (H).
Note that, as expected, QGP contribution dominates at high
masses, while hadronic contribution dominates at $\rho$
region and below. Although in the rates shown in Fig. \ref{rates} one
can see quite substantial contributions at low masses from $a_1$
decay, it is reduced by the small experimental acceptance in
 this region so dramatically that it falls well below the data.

  For check of consistency, we have
 compared our results with those of other approaches in 
 Fig.\ref{Mspectrum_conventional}(b). All three use ``conventional
 rates'' which are similar to ours (except the $a_1$ part) but very
 different dynamics of the evolution. Li-Ko-Brown approach is based on
 cascades
started from RQMD-based initial conditions\footnote{Note that
 it includes additionally the $\phi$
contribution 
ignored by us and others.}. 
  Srivastava $et.al.$ assumed Bjorken-scaling
longitudinal hydrodynamics, with solved hydro in the transverse
plane\footnote{Note that this curve is already corrected for the double counting 
which was present in the
  original paper, but their results are still somewhat larger.}.
We conclude that the agreement between 3 papers
is  reasonable in terms of the total yield, and even
very good in terms of the shape of the mass spectrum. Since all three
results
disagree with data,  
one may  conclude that it is not possible to
   explain CERES data
  by ``conventional sources''.

  In Fig.\ref{Mspectrum_conventional}(c) we also show our predictions
  for  160A GeV Pb+Au collisions. We compare them with new
  $preliminary$ CERES data reported at Quark Matter 96 \cite{CERES}.
Since the data are for non-central collisions, we have multiplied our
curve by the factor 1.7, the ratio of the $observed$ dilepton yield
for minimum bias sample $<n_{ch}>=260$ to that in central sample
 $<n_{ch}>\approx 400$. In this case the disagreement is not
 statistically as significant
as in the previous reaction, but still it seems that the measured shape
of the spectrum is different from the predicted one. 

\section{Dilepton yield from modified   $\rho,a_1$ mesons}
\label{sec_modified}

\begin{figure}[!h]
\includegraphics[width=13cm]{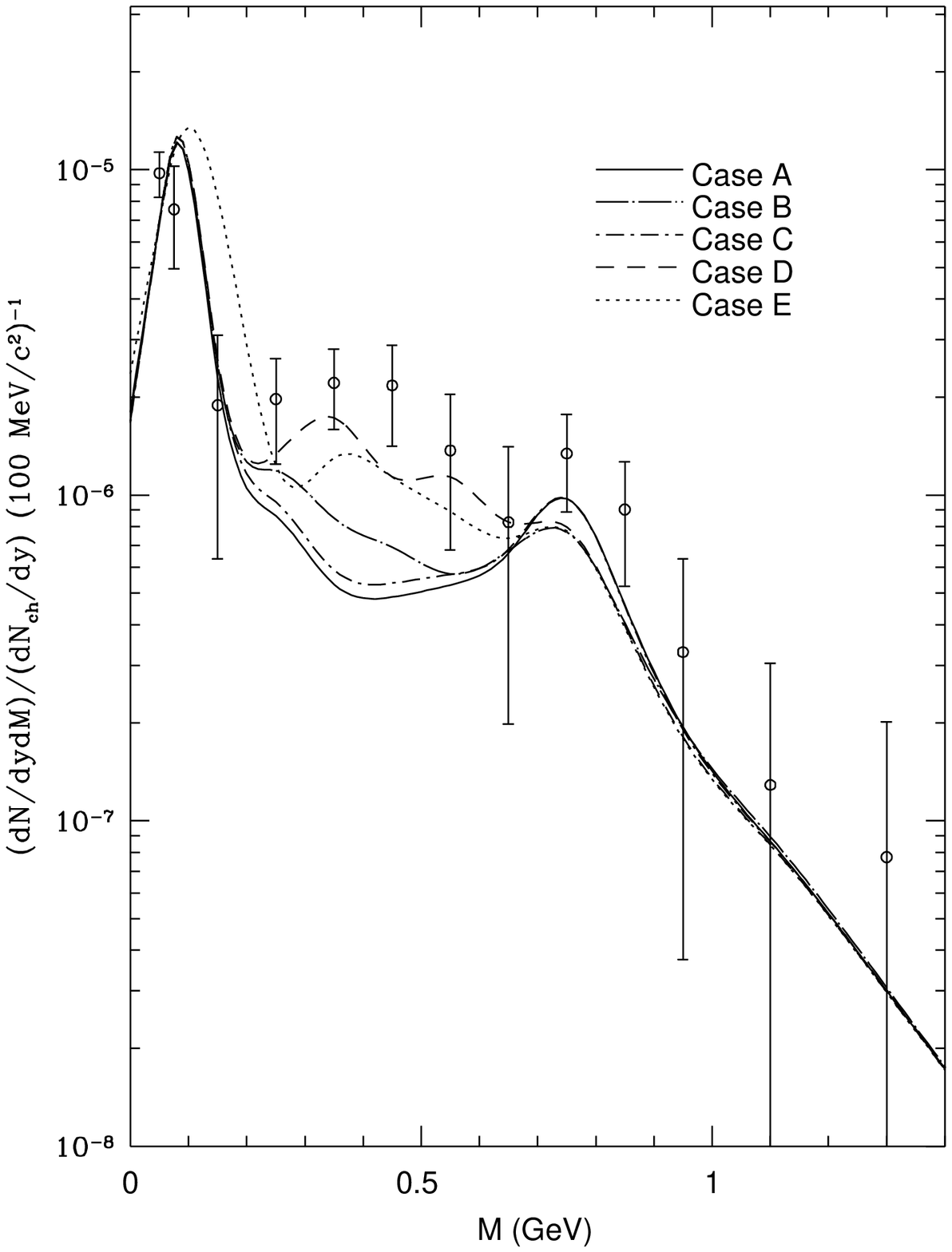}
\hspace{-10cm}
\includegraphics[width=4.5cm]{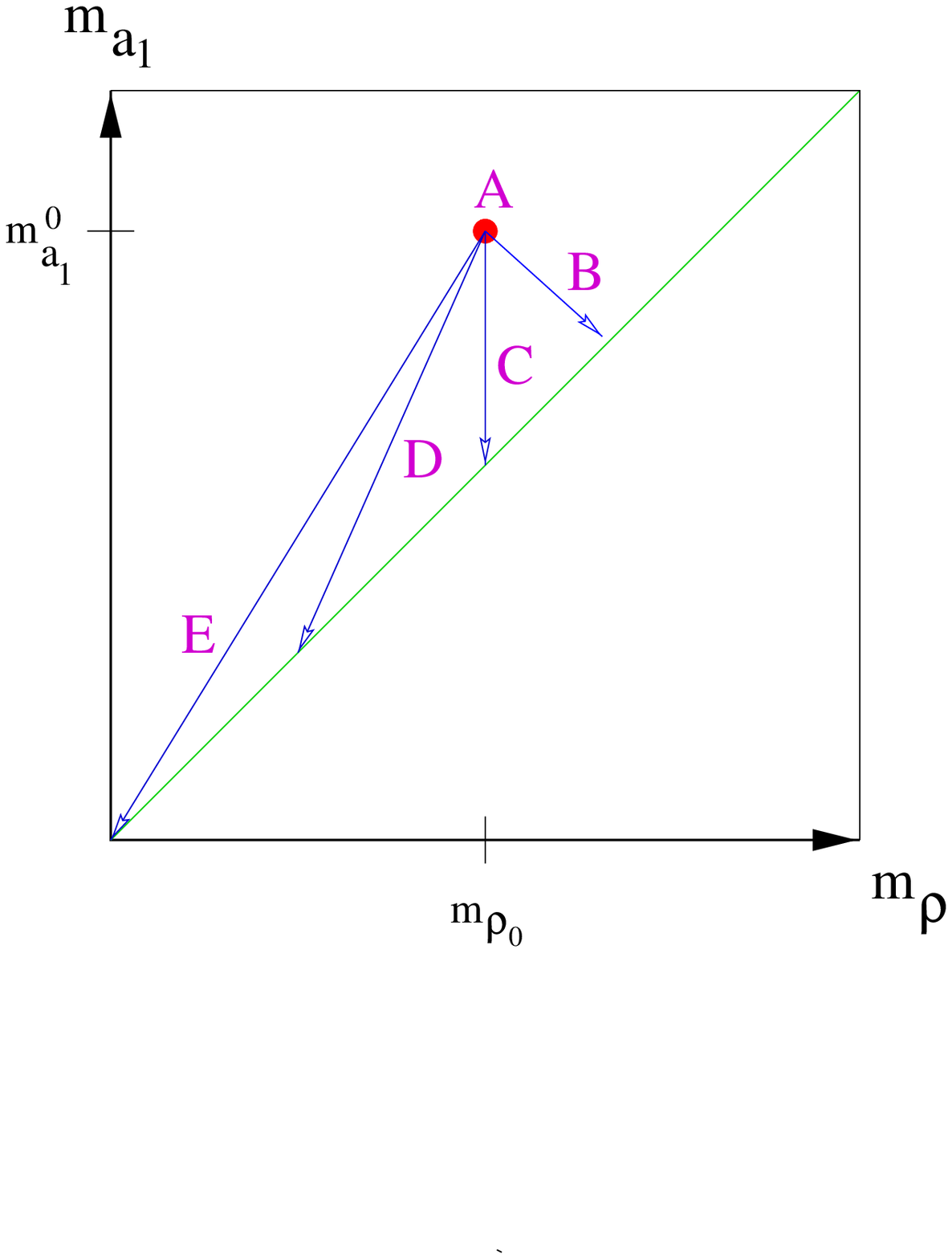}
\caption{\label{Mspectrum}
Dilepton yield in 200A GeV S+Au collision for different possible
scenario of chiral restoration in $m_\rho - m_{a_1}$ plane.}
\end{figure}
\begin{figure}[!h]
\begin{center}
\includegraphics[width=6cm]{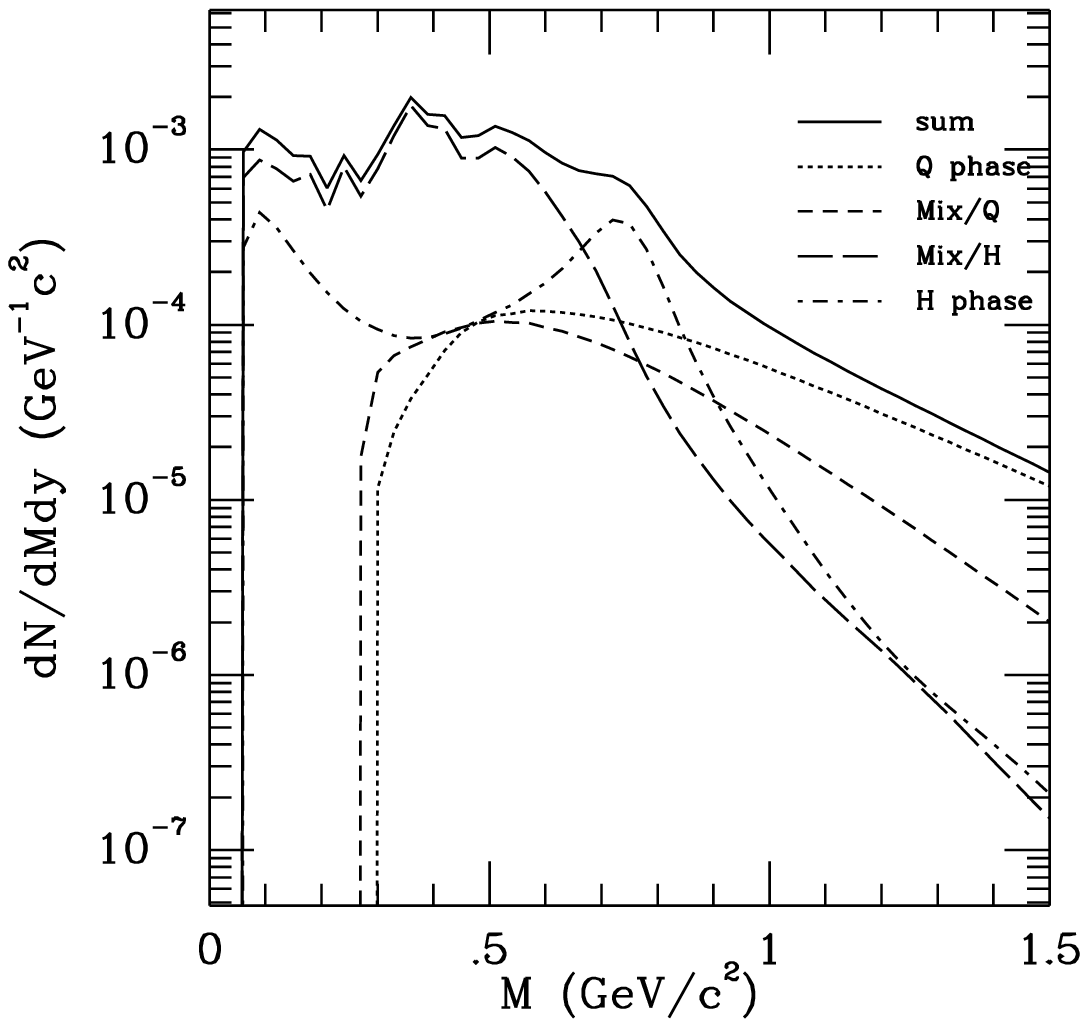}
\hspace{0.2cm}
\includegraphics[width=6cm]{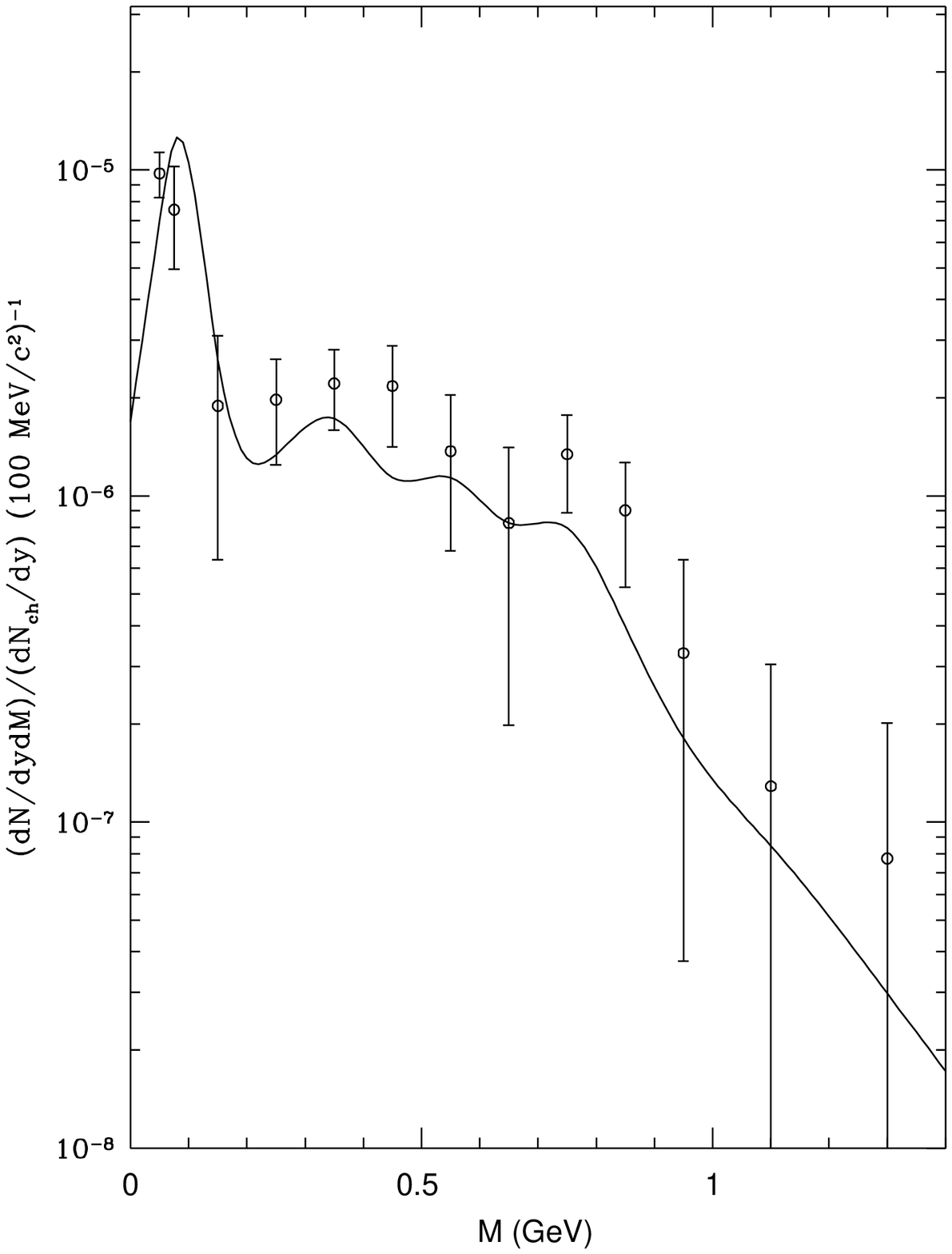}
\hspace{0.2cm}
\includegraphics[width=6cm]{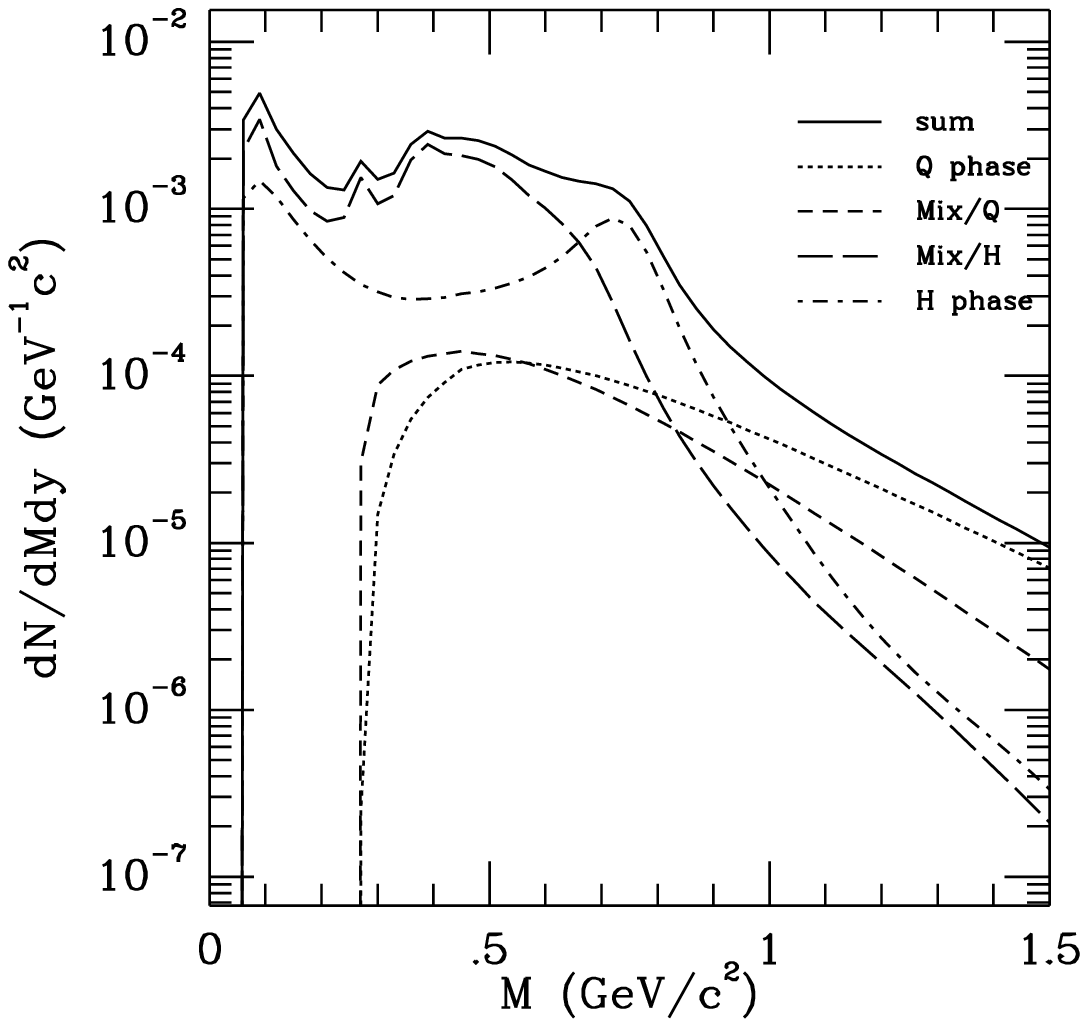}
\hspace{0.2cm}
\includegraphics[width=6cm]{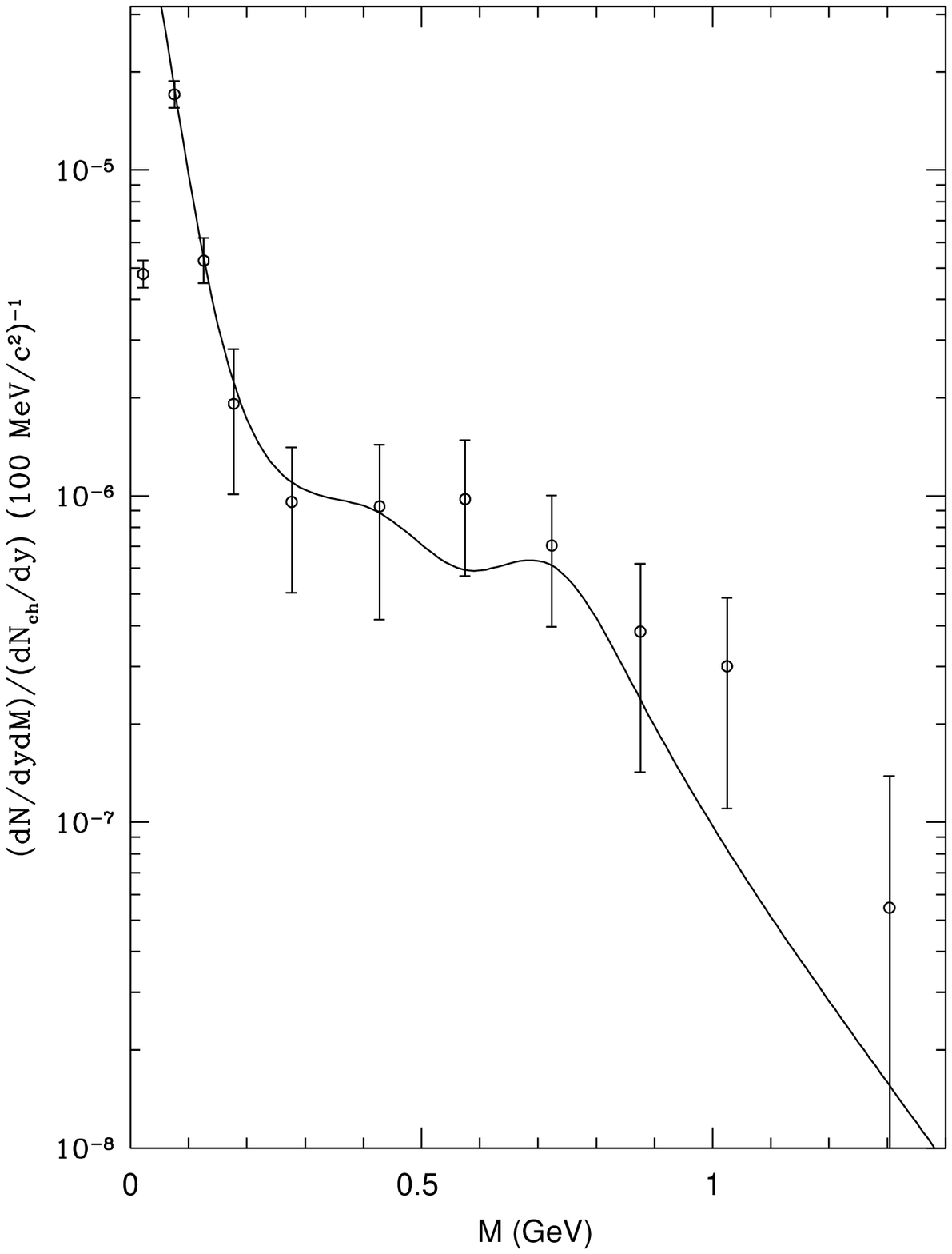}
\end{center}
\caption{\label{Mspectrum1}
Direct dilepton yields (without and with background)
for 200A GeV S+Au (top) and 160A GeV Pb+Au (bottom)
using mass scheme D.}
\end{figure}
  Dileptons, unlike secondary hadrons, 
are produced at relatively early stages of the collisions
One possible explanation of why the ``standard'' rates fail to reproduce
the observed excess of low-mass dileptons may be a long-debated idea:
hadronic properties 
 may be $modified$ in high density hadronic matter.

  We know definitely that the $nucleon$ mass is modified in nuclear
  matter, and shifts of the {\it vector meson masses} in it are debated for
  a long time, see e.g. \cite{rhoinnuclei,HSK_96}. 

 In a low-T hadronic gas (which may include
an admixture of baryons) one can relate modification of mesons (e.g. of $\rho$)
 to 
the $\pi\rho$ (and $N\rho$) forward scattering amplitudes
\cite{Shu_pot}.  This approach predicts certain
momentum-dependent optical potential, which can
be loosely interpreted as (relatively modest)  shift of $m_\rho$ downward.
Finite T/density QCD sum rules (see e.g.\cite{HK_94} and references therein)
relate hadronic properties to the quark condensate $<\bar q q>$, which
decreases toward chiral restoration transition. This reasoning has
culminated in the so called
  {\it Brown-Rho scaling idea}, according to which
all hadronic dimensional quantities get their scale from $<\bar q q>$:
 therefore $all$ masses are predicted to vanish  
at $T\rightarrow T_c$.

In the {\it instanton model} the chiral restoration is due to
transition from random instanton liquid to a gas of
instanton-anti-instanton molecules\cite{SSV_95}: the latter survive the phase
transition
and lead to new type of quark interactions unrelated to  $<\bar q q>$.
 Although results of available simulations \cite{SS_95} have not reported 
any definite conclusions about the rho meson mass, at quark level
 it was clearly demonstrated
that at $T>T_c$
an effective quark mass is substituted by ``effective energy'' (or
``chiral mass'') of comparable magnitude, which imply that hadronic
masses should $not$
vanish at $T \rightarrow T_c$.

 The matter however is by no means settled, and
 quite different suggestions about hadronic properties close to (or
 even above) $T_c$ can be
found in literature. For example, effective Lagrangians lead
to a prediction of $rising$ $m_\rho(T)$ \cite{Pisarski_95}, moving
it about half way toward the mass of its chiral partner $a_1$. 
Furthermore, at SPS about 1/10 of secondaries are baryons, and thus
finite baryon density effects should be added to the finite-T shifts
mentioned above. 

  Therefore, we have decided to proceed empirically, $assuming$ various   
scenarios of hadronic mass evolution in dense matter.
  A particular point we want to make in this paper
is that ``dropping  $m_\rho(T)$'' idea  is consistent with chiral
  symmetry restoration only if appropriate  modifications of its axial partner
 $m_{a_1}$ follow. Strict theoretical relation between the two were derived 
  via Weinberg-type sum rules
     \cite{KS_94}.
Possible scenarios of how chiral restoration may proceed are 
therefore shown in
the $m_\rho -   m_{a_1}$ plane in  Fig.\ref{Mspectrum}(a).
For example, path B corresponds to results of Pisarski \cite{Pisarski_95}
, while E corresponds to Brown-Rho scaling.
Anyway, both mesons should become
identical at $T_c$, so the path of thermal evolution should end up
at the diagonal   $m_\rho(T_c)=
  m_{a_1}(T_c)$.
   
  To test these ideas, we need a
simple  consistent model of $how$ a changing $\rho$ mass can change the
dilepton production rates discussed in section \ref{sec_rates}.
In this respect, the
standard expression for the pion form-factor (\ref{ff}) happen to be
quite misleading: it
includes $m_\rho$ both in numerator and in denominator, but simply to make it
T-dependent $everywhere$ would  in fact be a mistake. Recall the usual
reasoning why there is $m_\rho^4$ in numerator is that at $M=0$ the
form-factor should be F(0)=1. This statement should hold for 
the the $\rho$ contribution only if one assumes the so called
 {\it vector dominance}, demanding that the whole form-factor (and not
 just a part of it) is given by the rho pole. It is quite accurately
 satisfied
at T=0, but for T-dependent $m_\rho$ vector
 dominance has no reason to persist! 

  Actually the numerator of the form-factor contains a combination $m^2_\rho
  \Gamma_{e^+e^-}
\Gamma_\rho$, and in what follows we assume that both widths are
{\it T-independent}. Note that only assumption  $
\Gamma_{e^+e^-}=const(T)$
is actually important, because (i)  $m^2_\rho$ cancels when one
returns to non-relativistic form of Breit-Wigner; and (ii)
as we have argued above,  the value of the
$\Gamma_\rho$ is  nearly irrelevant for the rate anyway.

Let us  concentrate only on the $\rho$ part of the rate and
write the
dilepton production rate as (see appendix):
\begin{eqnarray}
\label{eq1}
\nonumber {dR\over d^4k} = && 
{-\alpha^2\over \pi^3}{m_{\rho_0}^4\over f_{\rho}^2M^2}
\left( 1+{2m_l^2\over M^2}\right) \left(
1-{4m_l^2\over M^2}\right)^{1\over 2}\\ && \times{Im \Pi\over \left(
M^2-\hat{m}_{\rho}^2\right)^2+\left( Im\Pi \right)^2}
\left( {1\over e^{\beta\omega}-1} \right)
\end{eqnarray}
where $\hat{m}_\rho^2 = m_{\rho_0}^2 + Re\Pi$ and $\Pi$ is the $\rho$
self-energy.  One can define the $M$-dependent width of $\rho$ by
the relation
$Im\Pi(T=0) \buildrel \rm def \over = - m_{\rho_0} \Gamma(M)$
which is generalized to non-zero T by
%
$Im\Pi(T,M) = - m_{\rho}(T,M) \Gamma(M)$
where we ignored the modification of the width with T.
Finally we can rewrite Eq.\ (\ref{eq1}) as
\begin{eqnarray}
\nonumber {dR\over d^4k} = && 
{2\alpha^2\over \pi^3}f_{\pi}^2{m_{\rho_0}^2\over M^2}
\left( 1+{2m_l^2\over M^2}\right) \left(
1-{4m_l^2\over M^2}\right)^{1\over 2}\\ && \times{m_{\rho}(T)\Gamma(M)\over \left(
M^2-\hat{m}_{\rho}^2\right)^2+m_{\rho}^2(T)\Gamma^2(M)}
\left( {1\over e^{\beta\omega}-1} \right)
\end{eqnarray}
where we have suppressed the $M$-dependence of $m_{\rho}$,
$\hat{m}_{\rho}$ and we used the relation $f_{\pi}^2 =
{m_{\rho_0}^2\over 2 f_{\rho}^2}$.

For simplicity we identify $\hat{m}_\rho(t)$ and $m_\rho(T)$, with
both given by
\begin{eqnarray}
{m(T)\over m_0} =
{a\over T/T_c - b} + r - {a\over 1 - b}
\end{eqnarray}
where $r$ is the value of ${m(t)\over m_0}$ when $T = T_c$, $a$
  characterizes the abruptness of the mass shift, while $b$ is chosen
  to make ${m(T)\over m_0} = 1$ at $T = T_f =$ 140 MeV.

\begin{figure}[!h]
\begin{center}
\includegraphics[width=6.cm]{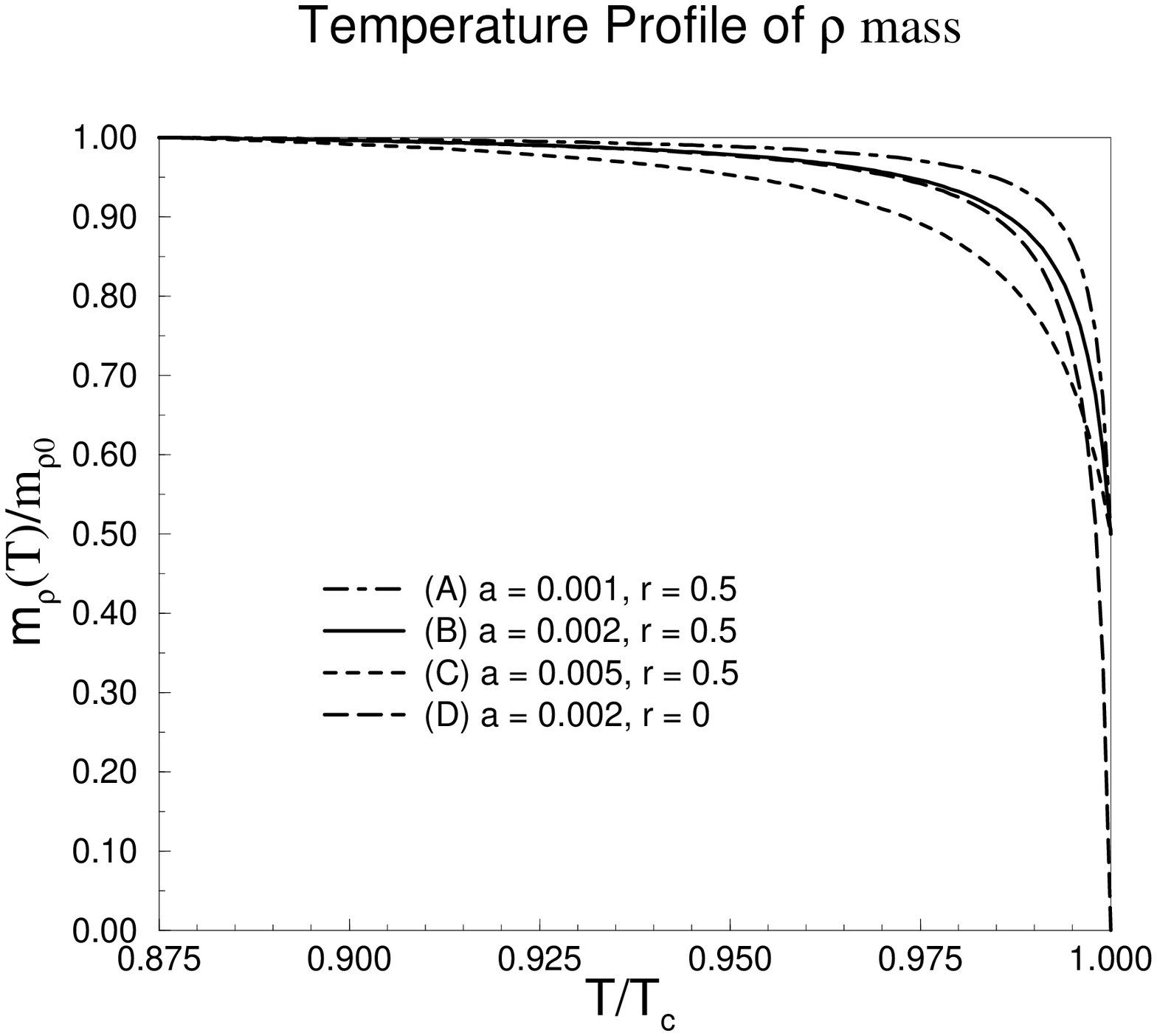}
\hspace{.2cm}
\includegraphics[width=6.cm]{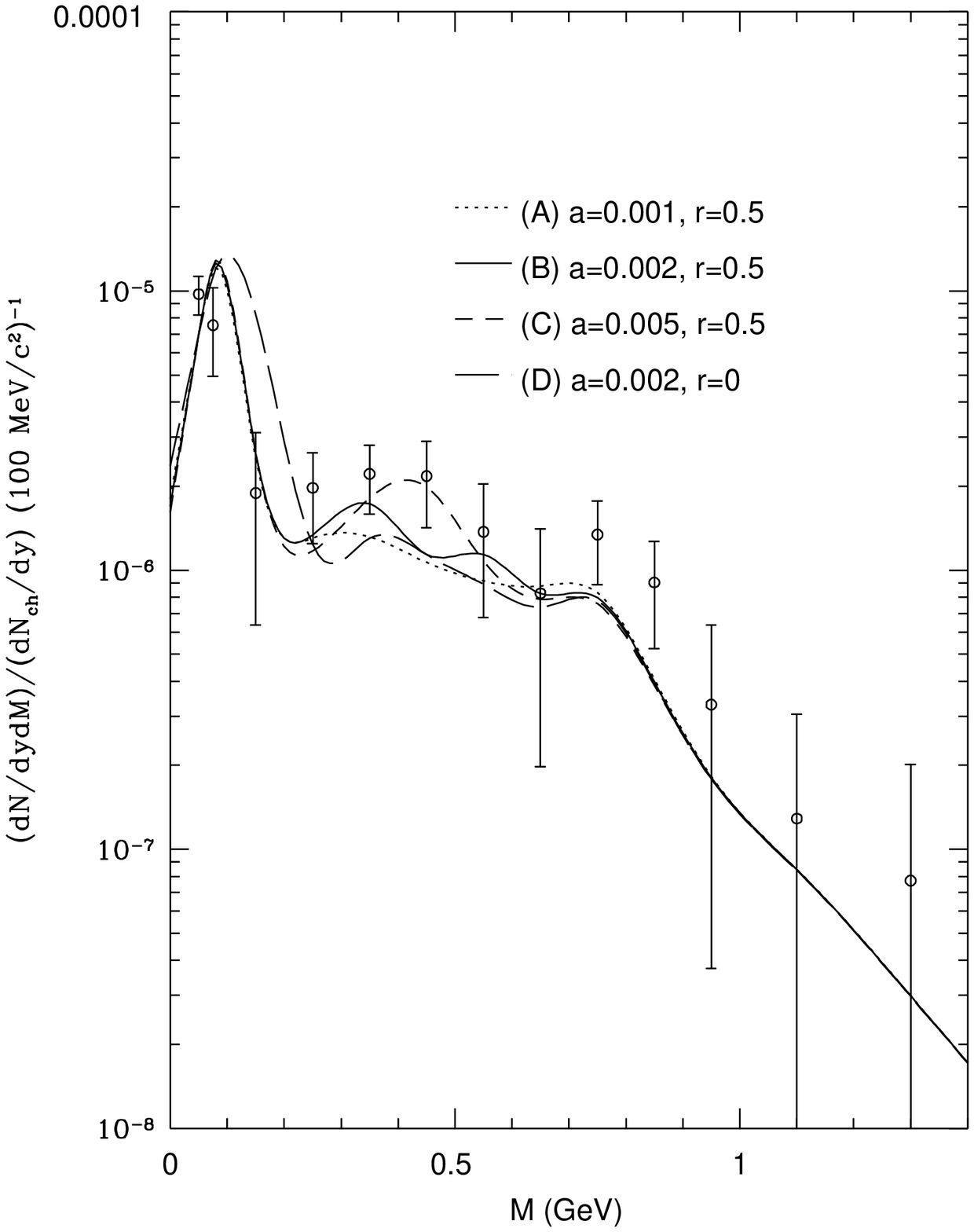}
\end{center}
\caption{\label{fig3}
Different temperature profiles for $m_{\rho}(T)$ (left)
 The corresponding dilepton spectra. (right)  For meaning of the
    parameters a and r, refer to equation (9)
}\end{figure}

   Now we can 
test whether a T-dependent  masses  can indeed
describe the CERES data in the lower mass region. Integrating over
space-time according to
hydro calculations described above we get the
dilepton mass spectra as shown in Fig.\ \ref{Mspectrum}.  
The variant D, with 
     $m_\rho(T_c)= m_{a_1}(T_c)\sim {1\over 2} m_\rho(0)$, does the best
     job, and Fig.\ref{Mspectrum1} we show contribution of separate 
stages in this scenario, for both 200 A GeV S+Au and 160A GeV Pb+Au.
It is the hadronic part of the mixed phase which 
is responsible for the observed excess at $M = 0.2 - 0.6$ GeV.
 Furthermore, depending on how exactly
 $m_{\rho}(T)$ goes to its limit at $T_c$, one can change the
 shape of the resulting mass spectrum.
This statement is demonstrated in Fig.\ref{fig3}, where (a) shows
several scenarios, with the resulting mass spectra in Fig.\ref{fig3}(b).
The case E in \ref{Mspectrum}, which corresponds to massless $\rho$
and $a_1$, generates more low-mass pairs than the hadronic cocktail.
Better data of $\eta$ production is needed to rule out this case.

  Of course, specific dynamical models lead to more complicated picture:
instead of simple universal  $m_{\rho}(T)$ for any $\rho$ meson, they
lead to specific modification of the whole dispersion curve. Clearly,
mesons which travel fast relative to matter are modified differently
from those which have zero velocity. Furthermore, since the
meson-meson and meson-baryon scattering
 is dominated by resonances, this dependence may even be
 non-monotonous.
Future high-statistics studies may look into those matters by
considering
mass spectrum at different rapidity and/or $p_t$ of the dileptons.

In conclusion: we have demonstrated that the CERES data can be fitted
pretty well over the whole experimental mass range by assuming a particular  
 $m_{\rho}(T)$, $without$ any change in the production mechanism or
 space-time
evolution.

\section{ Searching for the modified $a_1$ contribution}
\label{sec_a1}
  Great interest related with modified hadronic masses is explained by
  a possible relation between this phenomenon and the chiral phase
  transition. As it was repeatedly emphasized above, modification of
  $\rho$
imply certain modifications of $a_1$ as well. In this section we
discuss how this can be experimentally verified as well.
 
 The important role of $a_1$ for production of photons and dileptons was 
   discussed in  \cite{XSB,SX_where,Song_93,SKG_94}
 As we have shown in section \ref{sec_rates}, the main $a_1$
contribution is at low dilepton masses. However, 
 CERES has adopted cuts which makes its acceptance
very small at small  $M_{e^+e^-}$, and therefore, 
the $a_1$ contribution to its mass spectrum is not very significant.
  Those cuts were made
in order to get rid
of
hadronic background,  such as $\eta$ Dalitz decay. So, by simply
looking at small mass region one cannot find the   $a_1$ contribution.

Nevertheless, one can try to locate a kinematical ``window''
in which it may be better seen. We have found that the best range of
invariant dilepton 
masses is $M_{e^+e^-} \approx 300 MeV$. Above it
  hadronic background is relatively
small, but the contribution of $\rho$ channel and even of the quark
annihilation in the QGP phase exceed the  $a_1$ contribution. Below
this mass they are not so important, but soon the hadronic background
from ordinary Dalitz decays take over.

\begin{figure}[!h]
\begin{center}
\includegraphics[width=6.cm]{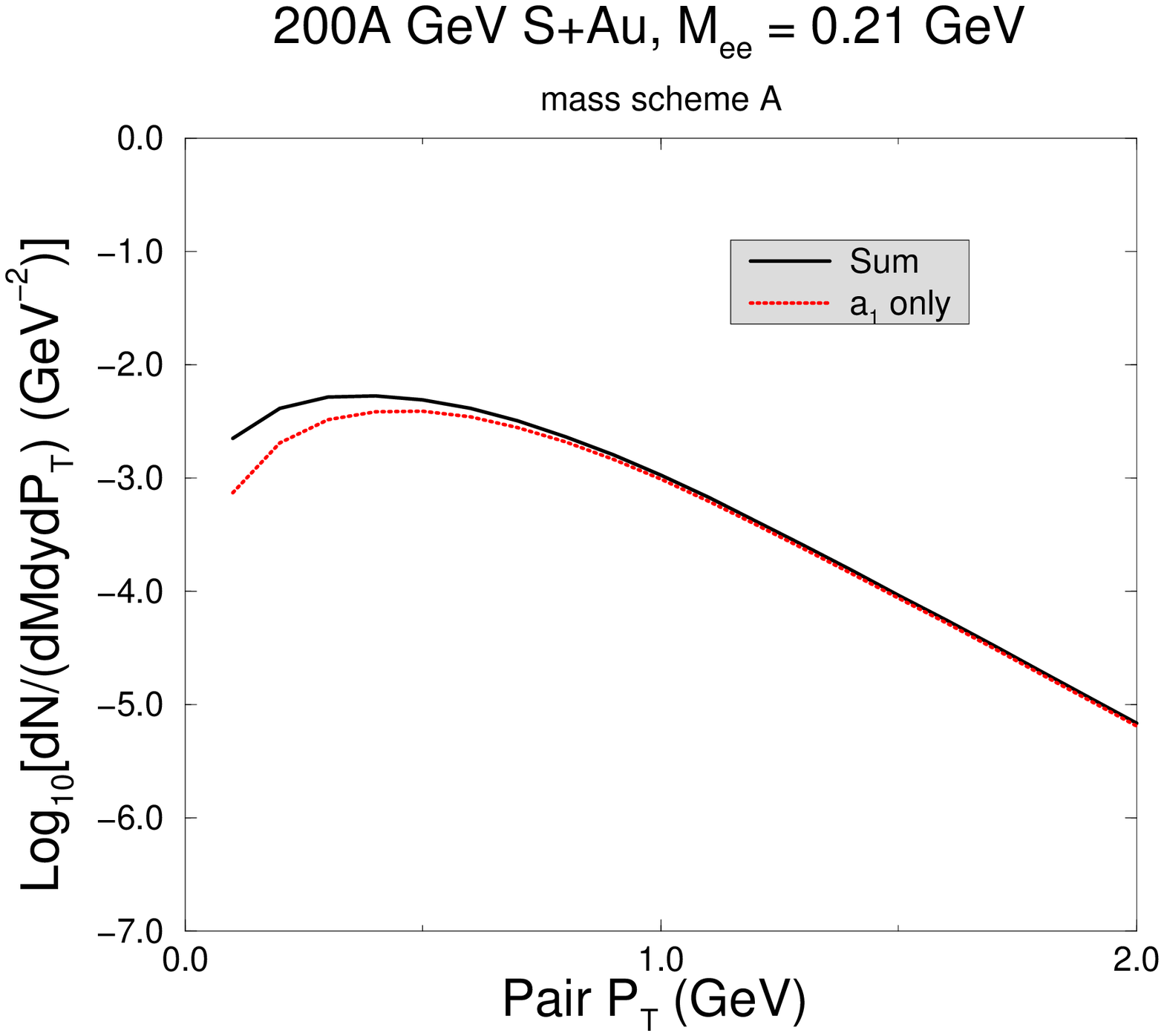}
\hspace{.2cm}
\includegraphics[width=6.cm]{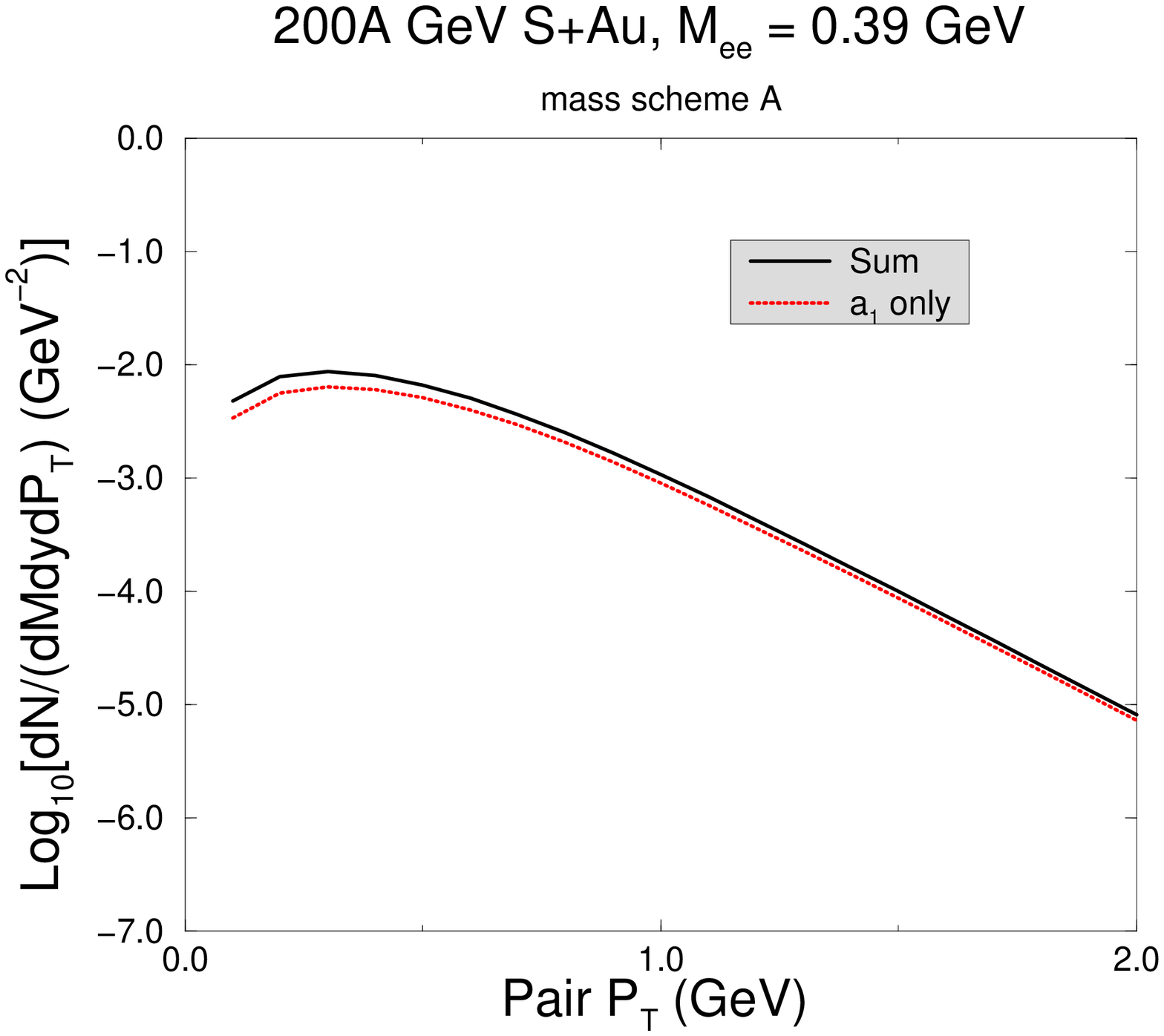}
\hspace{.2cm}
\includegraphics[width=6.cm]{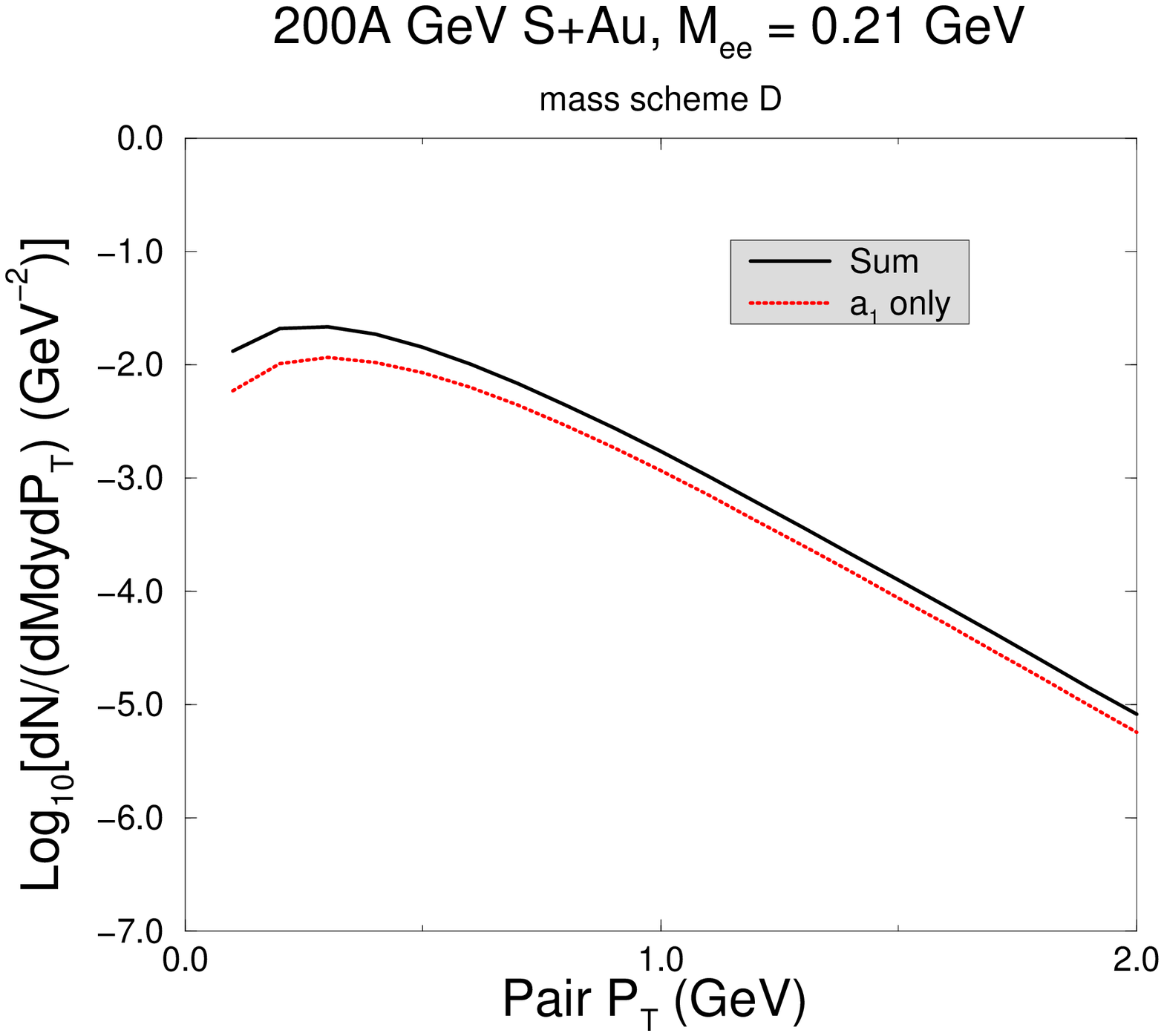}
\hspace{.2cm}
\includegraphics[width=6.cm]{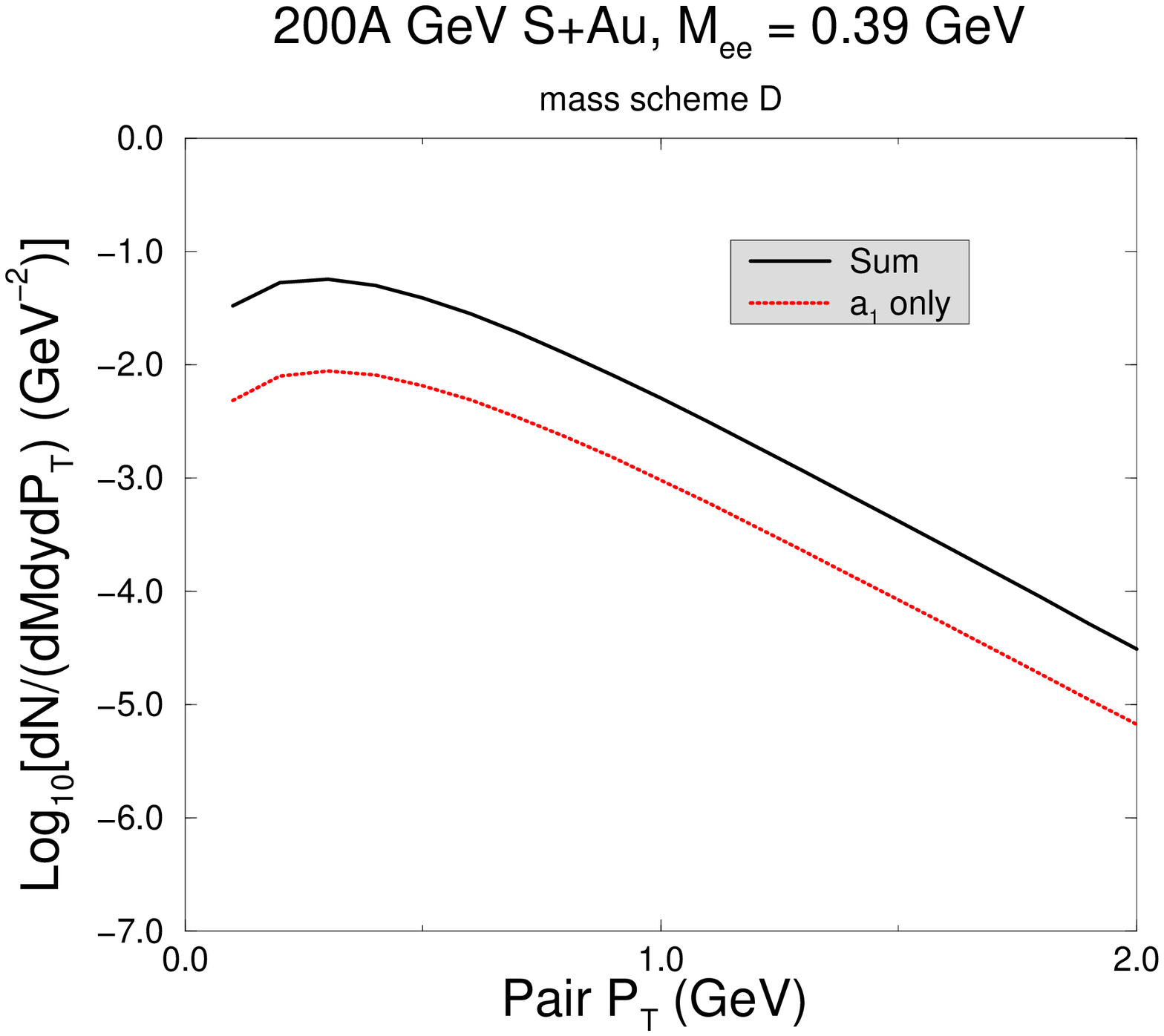}
\end{center}
\caption{\label{pt}
Dilepton $p_t$ spectrum for 200A GeV S+Au.  The top two figures are
for mass scheme A (i.e. unshifted masses) while the bottom two are for mass scheme D.
}\end{figure}

Furthermore, there is a particular $p_t$ dependence of the effect.
 Consider first the $unmodified$ $a_1$,
decaying into $\pi e^+ e^-$. For relatively small dilepton masses,
in the $a_1$ rest frame both the pion and dilepton get half of its
total mass. Including thermal motion and (rather large) $a_1$ width,
 one still finds a broad maximum in dilepton production rate 
 at dilepton energy
    about 600 MeV. 
Integrated over longitudinal momentum, one gets
    dilepton
$p_t$ distribution shown in Fig.\ref{pt}(a), with a maximum. It looks very different compared to the dileptons
coming from $\rho$ decay and having the usual thermal $p_t$ dependence.
   Now comes the main point: $if$ the $a_1$ mass shifts down, as
   expected from chiral restoration, then the
   wide peak
is absent and the shape is different: see  Fig.\ref{pt}(b).  

In summary, it is important to locate and to test properties of $a_1$
contribution. We have shown how sensitive dilepton production is to
its modification in dense matter. Although this task is not easy,
next generation of experiments (with accurate knowledge of hadronic
 backgrounds, measured dependence on the event centralities etc) can
 in principle do it.

\section{ Can one explain the dilepton excess by a long-lived fireball?} 
\label{sec_long}

\begin{figure}[!h]
\begin{center}
\includegraphics[width=12.cm]{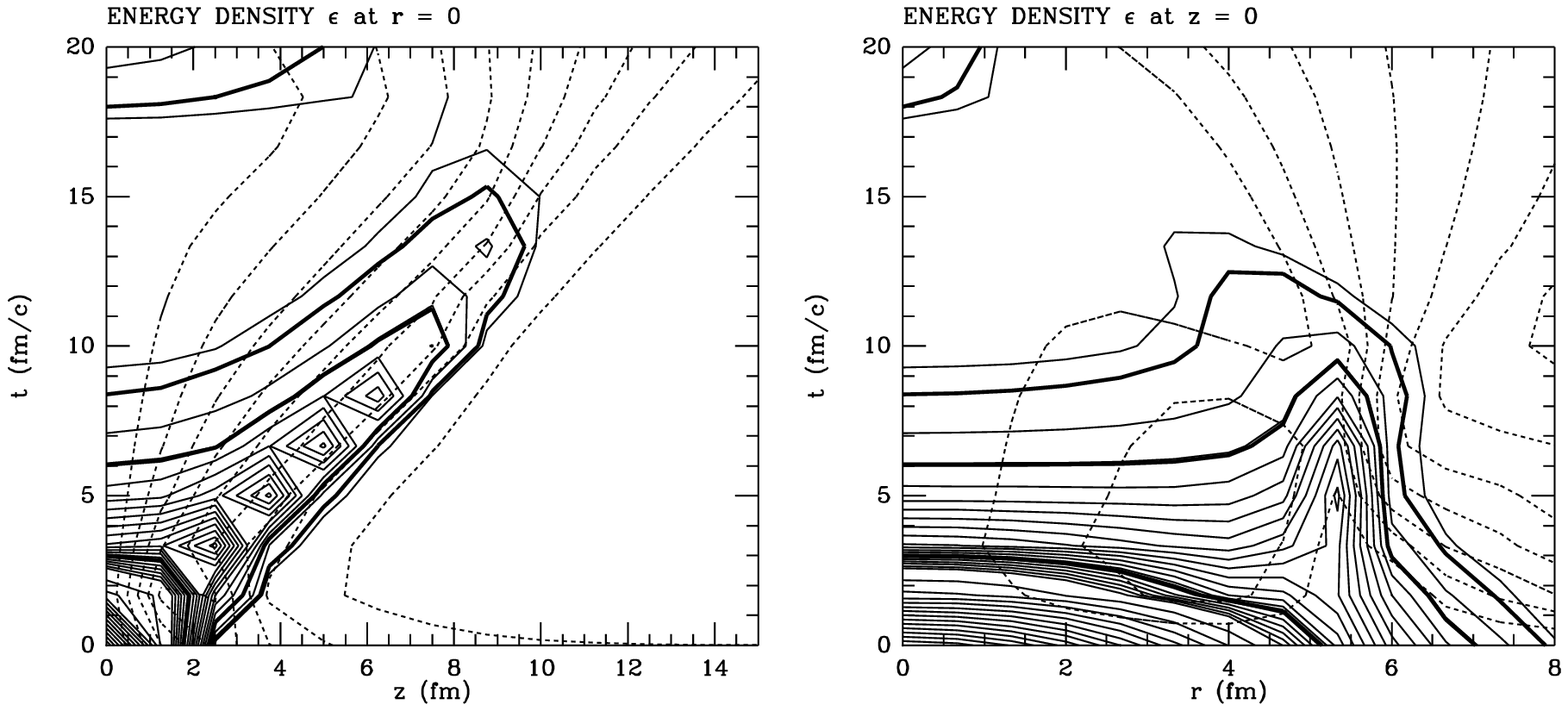}
\vspace{.2cm}
\includegraphics[width=12.cm]{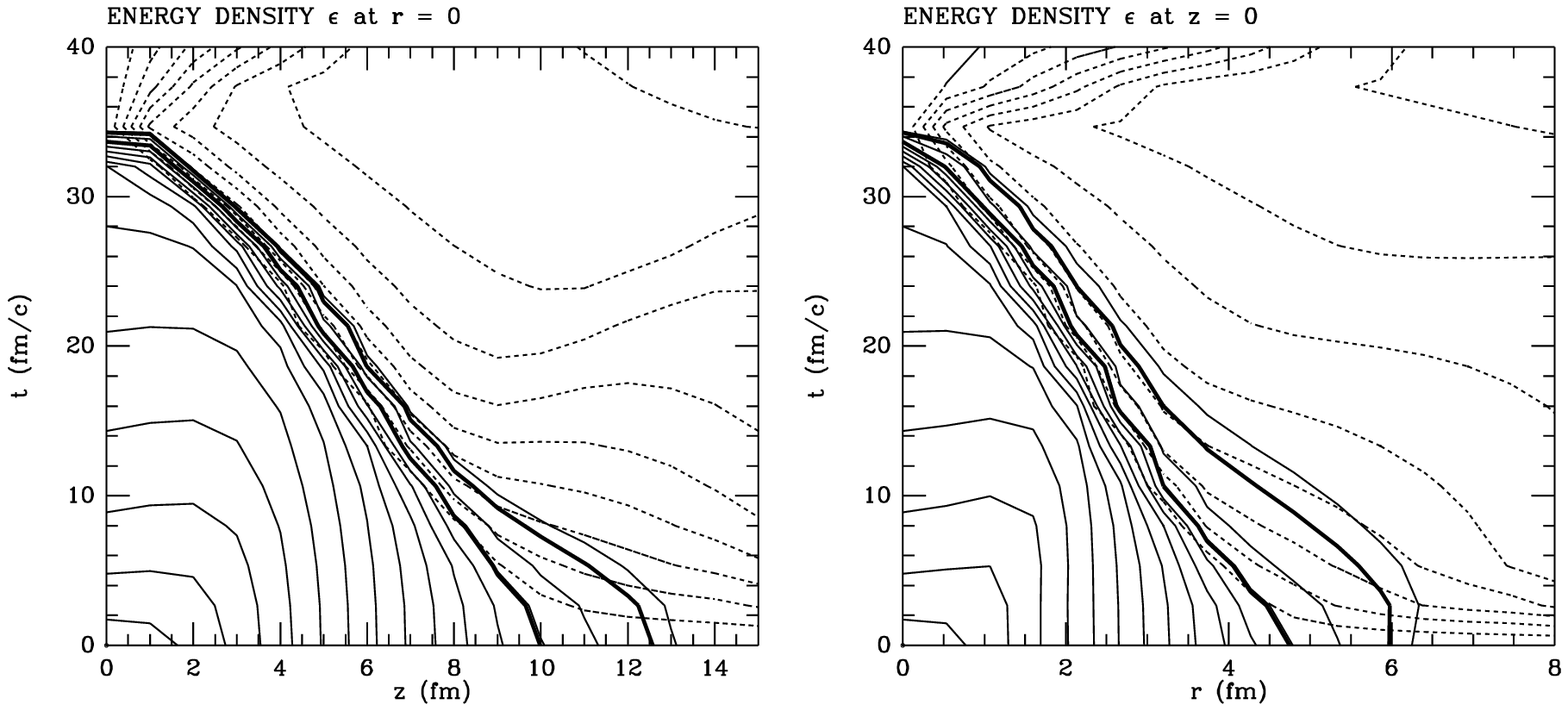}
\end{center}
\caption{\label{hydro}
Space-time evolutions for Standard Bjorken-like scenario (top)
   Long-lived fireball scenario (bottom)
}\end{figure}

  In our previous work \cite{HS_95} it was shown, that 
under certain conditions the ``softness'' of EOS
near the phase transition 
 can affect 
the $longitudinal$ expansion, so that even the global lifetime of the
excited system increases substantially. It happens in
a $window$ 
of collision energies such that the initial energy density is close to
the ``softest point'', which makes
secondary
acceleration of matter is  impossible.
Due to uncertain initial conditions, we do not exactly know to which collision
energies this window correspond, but estimates put it in the region
$E_{LAB}\sim 30 GeV/A$, between AGS and SPS.
Some examples of the solution we got at higher energies and at the
softest point are shown in 
  Fig.\ref{hydro}. One can see that the long-lived fireball
is slowly burning, with the lifetime reaching more than 30 fm/c!

It was further been proposed that one can test these unusual predictions
 experimentally by:
(i) Looking for the  maximal lifetime
 (or the minimum of  the HBT
   parameter $\lambda$) \cite{HS_95}; (ii)
 Looking for  the {\it minimum} of  the ``directed'' flow 
in the collision plane \cite{RG_96};
(iii) Looking for the nearly isotropic
distribution of  dileptons, produced in the long-lived fireball \cite{HS_95}.
  In connection with (i) it is very intriguing
 that the E802 AGS experiment
  reported preliminary studies of HBT  which indicate significant ($\sim 40
  \%$ )
 growth of lifetime for the most central Au Au collisions \cite{E802}.

  We now would like to check whether the long-lived fireball,
{\it if  present at SPS energies}, can enhance the dilepton production
to a degree necessary to explain the CERES data. For that we take the
initial conditions as shown in Table\ \ref{table1}, which indeed give the initial
energy density close to the softest point value. 

\begin{table}
\caption{Parameters for the two hydro models used}
\label{table1}
\begin{tabular}{lcc}
\hline
\hline
 &{\bf Bjorken-like} &{\bf Long-lived Fireball}\\
\hline
Initial energy density $\epsilon_i \rm(GeV/fm^3)$ & 9.0 & 1.5 \\
Initial temperature $T_i \rm(GeV)$ & 0.27 & 0.16 \\
Total energy in fireball $E_i \rm(GeV)$ & 830 & 930 \\
Initial long. velocity $v_{z_0}/c$ & 0.0 & 0.0 \\
Initial transverse vel. $v_{r_0}/c$ & 0.0 & 0.0 \\
Initial long. half-size $z_0 \rm(fm)$ & 1.1 & 8.0 \\
Initial transverse radius $r_0 \rm(fm)$ & 3.8 & 3.8 \\
Critical temperature $T_c \rm(GeV)$ & 0.16 & 0.16 \\
Freeze-out temperature $T_f \rm(GeV)$ & 0.14 & 0.14 \\
\hline
\hline
\end{tabular}
\end{table}

\begin{figure}[!h]
\begin{center}
\includegraphics[width=6.cm]{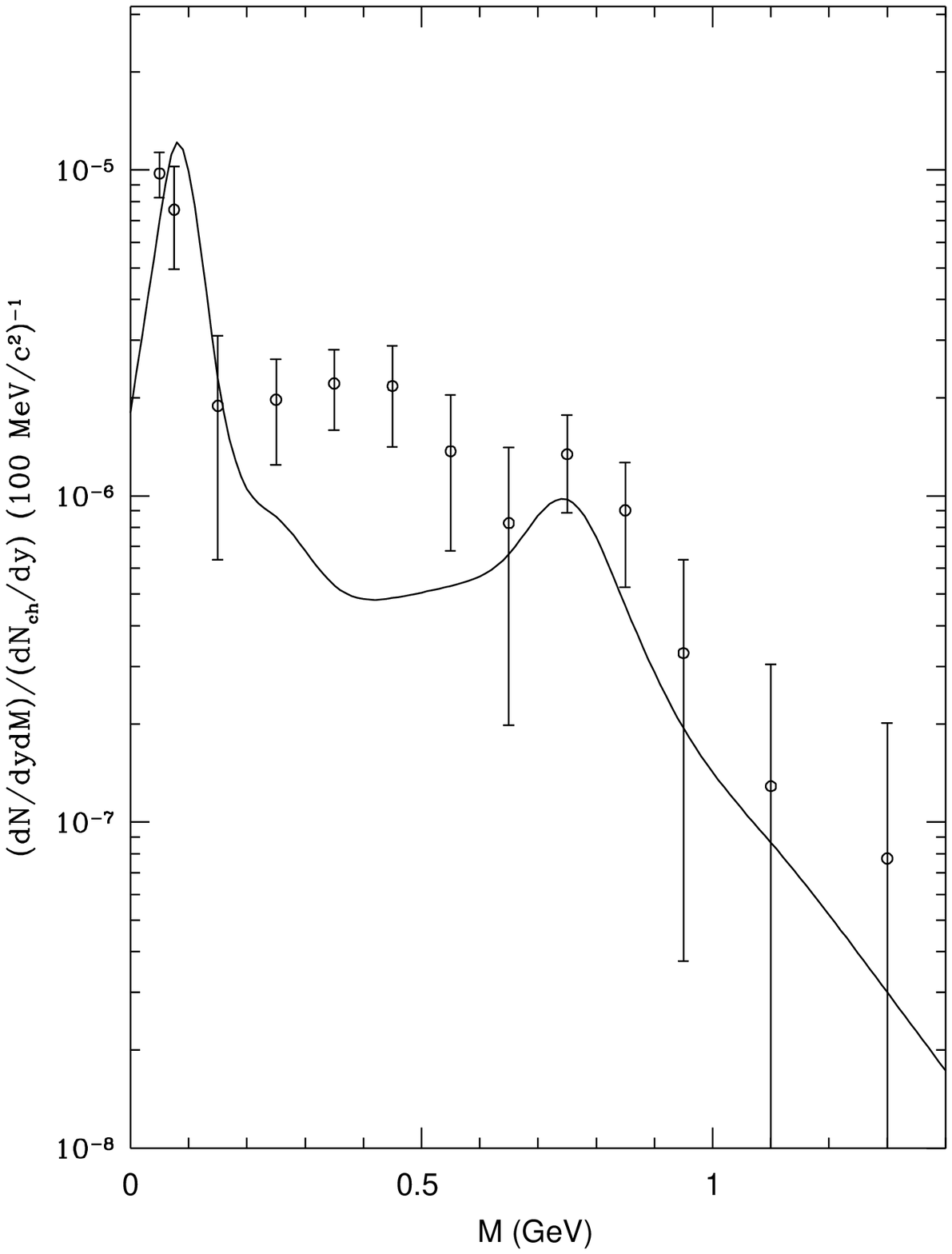}
\hspace{.2cm}
\includegraphics[width=6.cm]{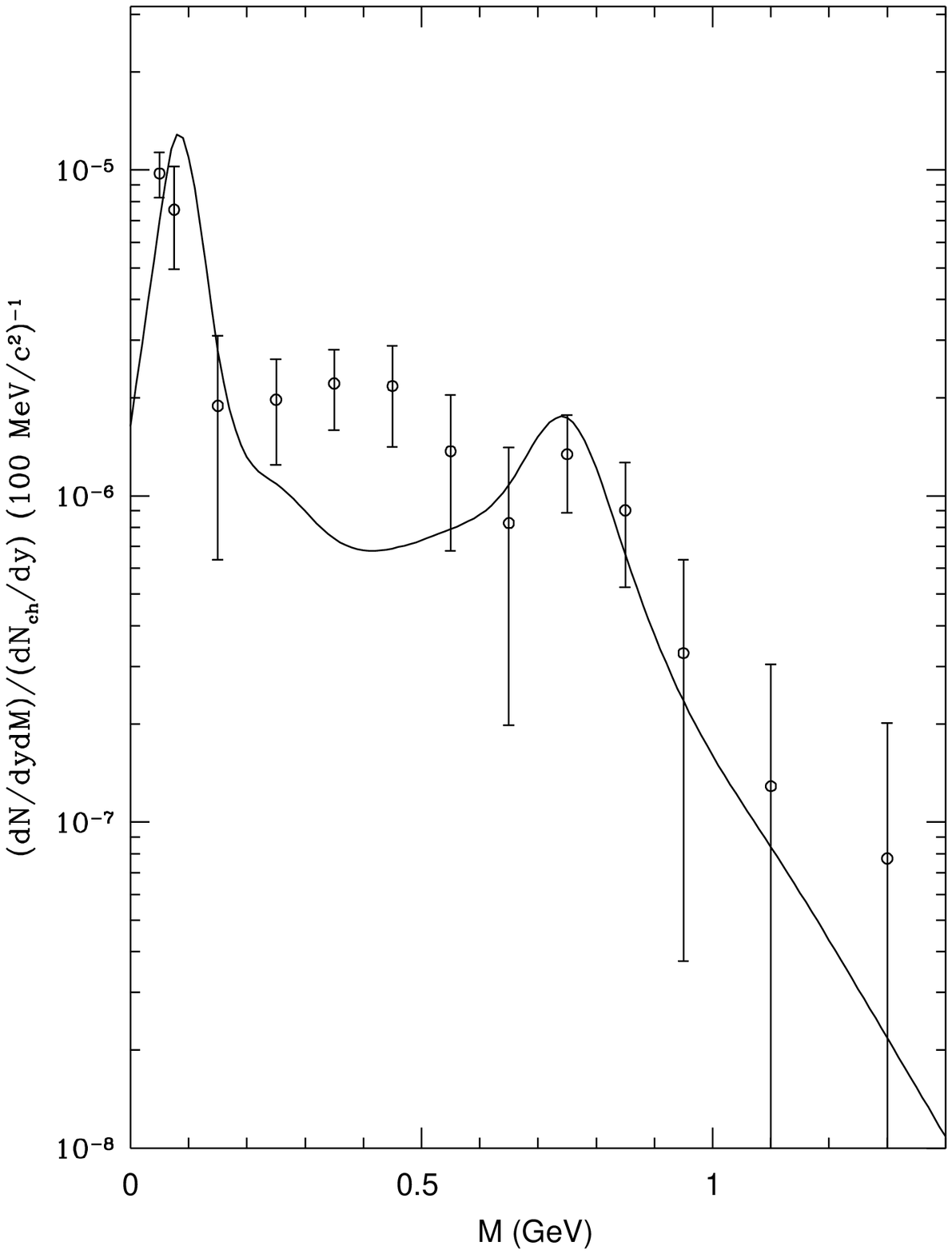}
\end{center}
\caption{\label{longlived}
200A GeV S+Au direct dilepton yields using unshifted masses for
the usual Bjorken-like expansion (left), and for
the long-lived fireball (right)
}\end{figure}

   The results however indicate that this scenario can neither be made
   consistent with rapidity distribution of secondaries, nor does it
   actually
result in larger production of dileptons. In Fig.\ref{longlived}
we compare two expansion scenarios with $unmodified$ masses. The
reason for this is as follows: longer lifetime is compensated
by smaller 3-volume. And also, the shape of the dilepton mass spectrum
is different from the observed one. 

\begin{figure}[!h]
\begin{center}
\includegraphics[width=6.cm]{ee8a.eps}
\hspace{.2cm}
\includegraphics[width=6.cm]{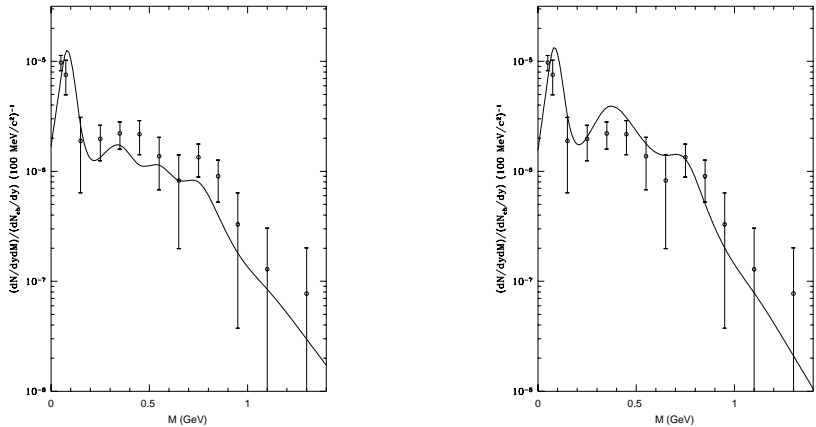}
\end{center}
\caption{\label{eefig}
200A GeV S+Au direct dilepton mass spectrum using the $\rho,a_1$ masses in scenario D for
Bjorken-like hydro (left) and for the
    long-lived fireball.(right)
}\end{figure}

Furthermore, if masses are modified (say according to variant D of Fig.\ref{Mspectrum}(a)),
both space-time pictures become compatible with data, see Fig.\ref{eefig}.

\section{Photon production}
\label{sec_photons}

  Let us briefly comment on the history of theoretical and experimental
  studies for photon production. As for dileptons, the radiation from QGP phase
  was
already calculated in \cite{Shu_78}. Reactions in a gas of
 $\pi,\rho$ mesons were considered in \cite{Kapusta_etal}, and
 importance
of $a_1\rightarrow \gamma \pi$ reactions was pointed out in
\cite{XSB}.  Following our dilepton calculations, we use the photon rate given
in \cite{SYZ_96}.

 Experimental observation of direct photons is an extremely difficult task.
 Unlike dileptons, we cannot tune the
  invariant mass and therefore for all photon momenta the background
  from hadronic decays (mostly $\pi^0,\eta$) dominates the signal of
  the
``direct'' photons. Therefore,  
   the  issue  is very accurate  measurements of these backgrounds
and inclusive photon spectra, with subsequent subtraction.
At the moment, only upper bounds on the direct photon cross section
has been given by the WA80 experiment \cite{WA80_seattle} \footnote{
  Earlier preliminary  WA80 data indicated a non-zero effect: its comparison
  with
theoretical expectations based on ``standard sources''
\cite{SX_where} lead to conclusion that those data significantly exceeded
the expectations. Later in \cite{ss} this
conclusion was disputed: larger photon yield (now consistent with
data)
 was obtained. The issue under debate
was
basically  how one should  normalize the hydrodynamical
initial conditions. Eventually, 
 the $preliminary$ data were withdrawn after
the reanalysis.}.

\begin{figure}[!h]
\begin{center}
\includegraphics[width=6.cm]{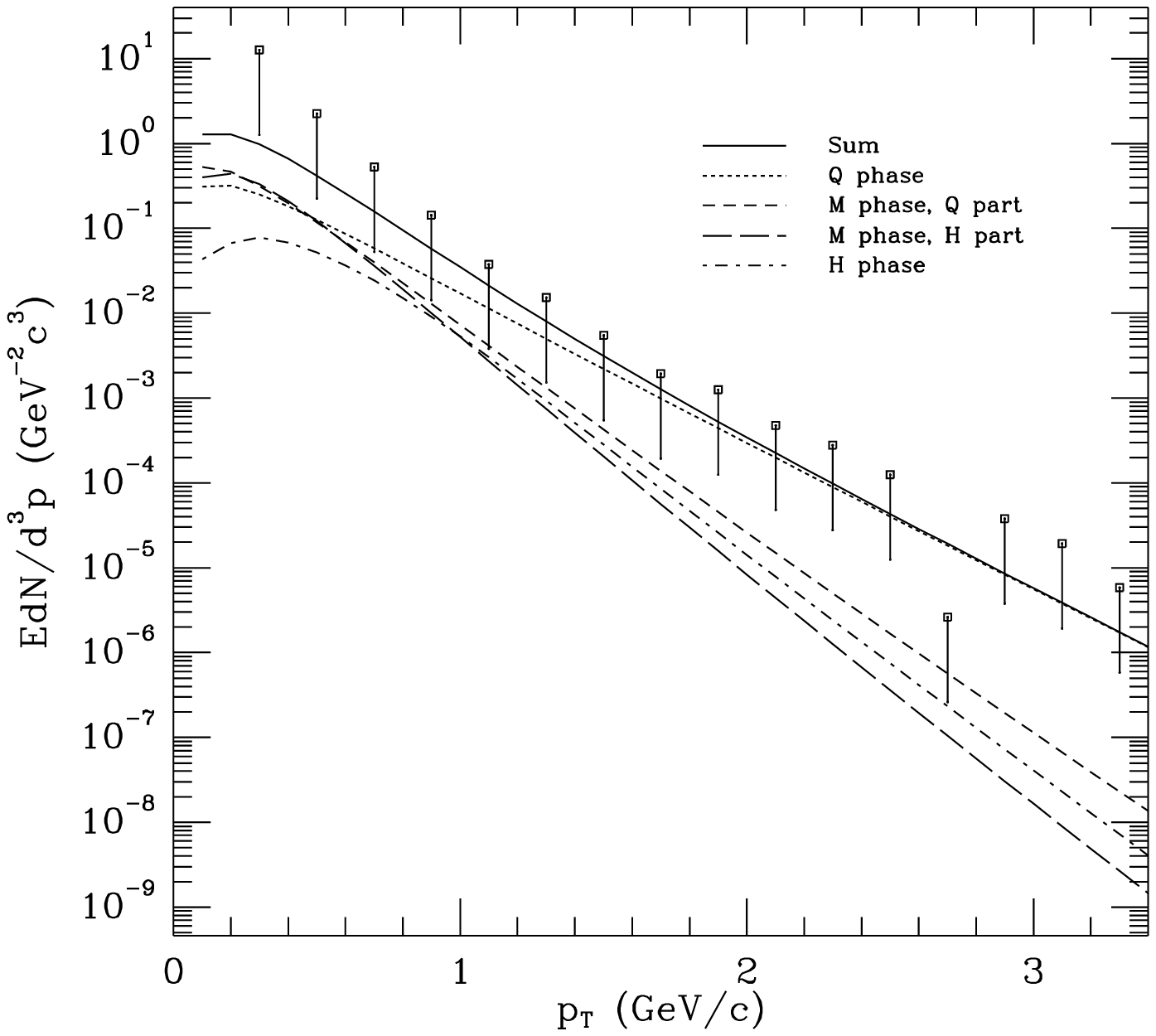}
\hspace{.2cm}
\includegraphics[width=6.cm]{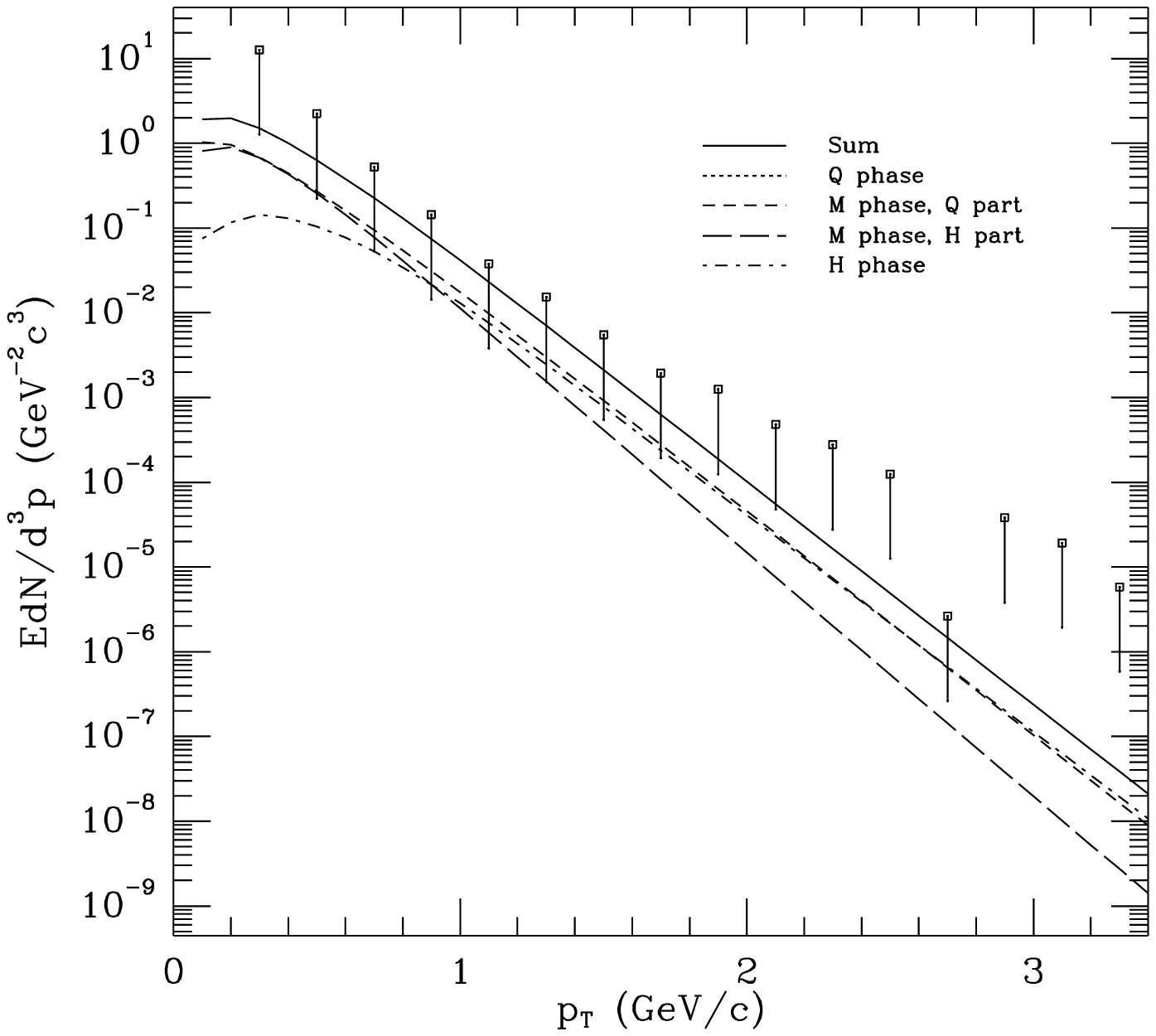}
\end{center}
\caption{\label{photon}
Direct photon production in 200A GeV S+Au compared with preliminary
WA80 upper bounds on direct photon production. Our predictions are
for the  Bjorken-like
expansion (left) with hot initial stage, and for the long-lived
fireball. (right)
}\end{figure}

 In Fig.\ref{photon} we show results of our calculations of the direct photon
 production. They correspond to unmodified hadronic parameters, and
are performed both for standard space-time scenario (a) and the
long-lived fireball (b). The main conclusions are: (i) The
theoretical  predictions are in both cases
$below$ the experimental bound in the whole region, although the
difference between them is not that large. (ii) The two scenarios
show significant difference at large $p_t\sim 3 GeV$, mostly due to
existence of relatively hot QGP in the first case. It would be very
important to pursue the issue further, in WA80 or elsewhere, and try
to observe radiation from QGP and hot hadronic gas.   

\section{Conclusions}
In summary, we have studied  dilepton/photon production
from highly excited hadronic matter produced in heavy ion collisions.

  Using the rates from the usual hadronic reactions with $vacuum$
  parameters and the usual hydro description of space-time picture,
we obtained results which are consistent with those of previous works but
cannot account for low-mass dilepton excess observed by CERES
experiment.
However, 
 if the $\rho$ mass is shifted to about 1/2 of its value around the
 critical temperature, the data can be explained. More detailed
observed distribution
(e.g. over dilepton mass and $p_t$) should provide better understanding
of whether this explanation of the observed dilepton excess is in fact correct.

 Furthermore, we have pointed out that chiral restoration demands 
that the mass of $a_1$ meson should be shifted together with the mass
of $\rho$
In order to get more convincing evidence that the excited matter is
indeed approaching chiral restoration, experimental observation
of the $a_1$-related component is essential. 
We discussed in which kinematical window one should look for it, and
evaluated
the magnitude of the effect.

  Among suggestions to explain the dilepton excess was a proposal that
the space-time picture of heavy
ion collisions can in fact be different, with possibly a longer-lived fireball.
However, although for specific  initial conditions
this
 long-lived scenario 
is hydrodynamically possible,   it
does $not$ significantly enhanced production of  dileptons because
longer lifetime is compensated by smaller spatial volume. Furthermore,
this scenario is incompatible with the observed spectra of secondaries. 

 Finally, all scenario considered lead to $photon$ production well
 below
current upper limit on direct photons from WA80. The results however
are sensitive to expansion scenario. The QGP component in particular
becomes dominant around $p_t\sim 3 GeV$.

\vskip 1cm
{\bf Acknowledgments} 
We'd like to thank G.Brown, I.Zahed, M.Prakash and  J.Steele for stimulating discussions.
This work is  supported by the 
US Department
 of Energy under Grant No. DE-FG02-88ER40388.

\eject
\section*{Appendix: Thermal Dilepton Rate}
It can be shown \cite{vdm} that the thermal rate of dilepton
production in a hot hadron gas is:
\begin{eqnarray}
\nonumber {dR\over d^4k} = &&{2e^2\over (2\pi)^6} {1\over (k^2)^2} Im
\Pi^R_{\mu\nu}(k) {1\over e^{\beta\omega}-1}\int {d^3p_+\over E_+}
\int {d^3p_-\over E_-} \delta^4(p_+ + p_- - k)\\ && \left[ p_+^\mu p_-^\nu
+ p_+^\nu p_-^\mu - g^{\mu\nu}(p_+\cdot p_- + m_l^2) \right]
\end{eqnarray}

Using vector dominance to relate the imaginary part of the photon
self-energy $\Pi_{\mu\nu}^R$ and the imaginary part of the $\rho$
propagator $\Delta^R_{\mu\nu}$ \cite{vdm}, then integrating over $p_+$ and $p_-$, we get
\begin{eqnarray}
\nonumber {dR\over d^4k} = {\alpha^2\over 3\pi^3}{m_{\rho}^4\over f_{\rho}^2M^4}
&& \left( 1+{2m_l^2\over M^2}\right) \left(
1-{4m_l^2\over M^2}\right)^{1\over 2}\left(
k^{\mu}k^{\nu}-M^2g^{\mu\nu}\right)\\ &&
\times Im \Delta^R_{\mu\nu}(k) {1\over e^{\beta\omega}-1}
\end{eqnarray}

This can be manipulated further into the form \cite{koch}\cite{vdm}
\begin{eqnarray}
\nonumber {dR\over d^4k} = && {2\omega dR\over dM^2d^3{\bf k}} =
{-\alpha^2\over \pi^3}{m_{\rho}^4\over f_{\rho}^2M^2}
\left( 1+{2m_l^2\over M^2}\right) \left(
1-{4m_l^2\over M^2}\right)^{1\over 2}\\ && \times{Im \Pi\over \left(
M^2-\hat{m}_{\rho}^2\right)^2+\left( Im\Pi \right)^2}
\left( {1\over e^{\beta\omega}-1} \right)
\end{eqnarray}
where $\hat{m}_{\rho}^2 \equiv m_{\rho}^2 + Re \Pi$ and $\Pi$ is now
the transverse or longitudinal part of the $\rho$ self-energy.  
We'd now demonstrate that the above rate can be written in terms
of thermal $\rho$ decay rates.
we are
interested in the invariant mass distribution (assuming $\Pi = \Pi(M)$),
\begin{eqnarray}
\nonumber {dR\over dM^2} =
{-\alpha^2\over 2\pi^3}{m_{\rho}^4\over f_{\rho}^2M^2}
\left( 1+{2m_l^2\over M^2}\right) \left(
1-{4m_l^2\over M^2}\right)^{1\over 2} && {Im \Pi\over \left(
M^2-\hat{m}_{\rho}^2\right)^2+\left( Im\Pi \right)^2}\\ &&
\times\int{d^3{\bf k}\over \omega}{1\over e^{\beta\omega}-1} \label{dRdM2}
\end{eqnarray}

which in the Boltzmann approximation reduces to:
\begin{eqnarray}
\nonumber {dR\over dM^2} =
{-2\alpha^2\over \pi^2}{m_{\rho}^4\over f_{\rho}^2M^2}
MTK_1\left({M\over T}\right) &&
\left( 1+{2m_l^2\over M^2}\right) \left(
1-{4m_l^2\over M^2}\right)^{1\over 2}\\ && \times{Im \Pi\over \left(
M^2-\hat{m}_{\rho}^2\right)^2+\left( Im\Pi \right)^2}
\end{eqnarray}
Following the notations of \cite{mt_scaling}
 we write
the dilepton rate as:
\begin{eqnarray}
{dR\over dM^2} =
{\alpha^2\over 24\pi^3}
MTK_1\left({M\over T}\right)
\left( 1+{2m_l^2\over M^2}\right) \left(
1-{4m_l^2\over M^2}\right)^{1\over 2}
F_{eff}(M)
\end{eqnarray}
where we define
\begin{eqnarray}
F_{eff}(M) \equiv
{m_{\rho}^4\left( -{48\pi\over f_{\rho}^2M^2}Im \Pi\right)\over \left(
M^2-\hat{m}_{\rho}^2\right)^2+\left( Im\Pi \right)^2} \label{F_eff}
\end{eqnarray}
To one loop\cite{vdm} (i.e. consider only $\pi - \pi$ annihilation), we have
\begin{eqnarray}
Im \Pi = -{f_{\rho}^2\over 48\pi} M^2\left(
1-{4m_{\pi}^2\over M^2} \right)^{3\over 2}
\theta\left( M^2-4m_{\pi}^2 \right)
\end{eqnarray}
Thus we recover the usual form of the form-factor,
\begin{eqnarray}
\nonumber F_{eff}(M)(1\ loop) = &&
{m_{\rho}^4\over \left(
M^2-\hat{m}_{\rho}^2\right)^2+\left( m_{\rho}\Gamma_{\rho\ total}
\right)^2} \left(
1-{4m_{\pi}^2\over M^2} \right)^{3\over 2}\\ &&
\times\theta\left( M^2-4m_{\pi}^2 \right)
\end{eqnarray}
where we've defined $Im \Pi \equiv - m_{\rho}\Gamma_{\rho\ total}$ as
usual.

If we include other interactions besides $\pi - \pi$ annihilation, in
general $Im \Pi$ (or $\Gamma_{\rho\ total}$) will increase, which
increases the width of $F_{eff}$ in Eq.\ (\ref{F_eff}).  However, the
maximum of $F_{eff}$ is given by
\begin{eqnarray}
F_{eff\ max} = F_{eff}(\hat{m}_{\rho}) = -{48\pi m_{\rho}^4\over
f_{\rho}^2\hat{m}_{\rho}^2 Im \Pi}
\end{eqnarray}
where we assumed that $\hat{m}_{\rho}$ and ${Im \Pi\over M^2}$ are
slowly varying functions of $M$.  Thus we see that the peak of
$F_{eff}$ decreases with increasing $Im \Pi$.  This contradicts the
conclusions in \cite{mt_scaling} where increases in {\it both} the
width and height of $F_{eff}$ were reported.  The discrepancy can
partly be explained by double counting of dilepton rates in
\cite{gale_lichard}, which is the basis for $F_{eff}$ in
\cite{mt_scaling}.  To see this, we rewrite the dilepton rate
(\ref{dRdM2}) as
\begin{eqnarray}
{dR\over dM^2} =
{3\over 8\pi^4}
{m_{\rho}^2\Gamma_{\rho\ total}(M)\Gamma_{\rho\rightarrow ll}(M)\over \left(
M^2-\hat{m}_{\rho}^2\right)^2+\left( m_{\rho}\Gamma_{\rho\ total} \right)^2}
\int{d^3{\bf k}\over \omega}{1\over e^{\beta\omega}-1} \label{breit_wigner}
\end{eqnarray}
where the decay widths are given by\cite{koch}\footnote{Note that $\Gamma_{\rho\rightarrow\pi\pi}$ in
(2.18) of \cite{koch} has a factor of 16 instead of 48.}

\begin{eqnarray}
\Gamma_{\rho\rightarrow ll}(M) =
{\alpha^2\over f_{\rho}^2/4\pi}\left(
{m_{\rho}^3\over 3M^2}\right)
\left( 1+{2m_l^2\over M^2}\right) \left(
1-{4m_l^2\over M^2}\right)^{1\over 2}
\end{eqnarray}
\begin{eqnarray}
\nonumber \Gamma_{\rho\ total}(M) \simeq
\Gamma_{\rho\rightarrow\pi\pi}(M) && =
{f_{\rho}^2\over 48\pi}{M^2\over m_{\rho}}
\left( 1-{4m_{\pi}^2\over M^2} \right)^{3\over 2}
\theta\left( M^2-4m_{\pi}^2 \right)\\ &&
 = -{Im \Pi\over m_{\rho}}
\end{eqnarray}

We note that (\ref{breit_wigner}) is essentially the same as the 
thermal $\rho$ decay
rate (2.9) of \cite{gale_lichard}, which was {\it added} to the $\pi -
\pi$ annihilation and other reactions in \cite{gale_lichard}.  
But Eq.\ (\ref{breit_wigner})
is simply a statement of the fact that the dilepton rate can be
interpreted {\it either} as coming from annihilation of pions {\it or}
as coming from the decay of thermal $\rho$'s, the invariant mass
distribution of which is given by the Breit-Wigner form (see also
\cite{weldon}\cite{koch}).  We cannot,
however, include {\it both} the thermal decay of $\rho$'s and $\pi -
\pi$ annihilation.  Doing so, as in \cite{gale_lichard} amounts to
double-counting of the dilepton rates coming from the $\rho$ channel.
\footnote{In \cite{gale2} this double-counting has been noted and corrected}
\vfill
\eject


\par

\end{document}